\documentclass[%
 reprint,
bibnotes,
 amsmath,amssymb,
 aps,
prb,superscriptaddress
]{revtex4-2}

\usepackage{graphicx}

\usepackage{textcomp}
    \usepackage{amsthm}
    \usepackage{amsfonts}
    \usepackage{amssymb}
    \usepackage{epstopdf}
    \usepackage{cleveref}

\begin{document}
 \preprint{APS/123-QED}
\title{Design of Sn-doped cadmium chalcogenide based monolayers for valleytronics properties}

\author{Sutapa Chattopadhyay and Anjali Kshirsagar}
\affiliation{Department of Physics, Savitribai Phule Pune University,  Pune 411007,  India}

\vspace{10pt}

\date{\today}

\begin{abstract}
Valleytronics, that uses 
the valley index or valley pseudospin to encode information, has emerged as an interesting field of research in 
two-dimensional (2D) systems with promising device applications. Spin-orbit coupling (SOC) and inversion 
symmetry breaking leads to spin-splitting of bands near the energy extrema (valleys). 
 In order to find a new 2D material useful for 
valleytronics, we have designed hexagonal planar monolayers of cadmium chalcogenides (CdX, X = S, Se, Te) 
from the (111) surface of bulk CdX zinc blende structure. Band structure study  reveals  valence band local maxima at symmetry point K 
and its time reversal conjugate point K$\textquotesingle$.  Application of SOC initiates spin-splitting in the valleys 
that lifts the energy degeneracy and shows strong valley-spin coupling character. We have substituted two Cd atoms in the planar monolayers by Sn atoms which 
increases the spin-splitting significantly.
The structural, dynamic, mechanical  and thermal stability of all the monolayers
 has been confirmed. Values of formation energies indicate that it may be feasible to synthesize the Sn-doped CdSe and CdTe monolayers using bottom-up
 approach.
Zeeman-type spin-splitting is observed in the valley region and Rashba spin-splitting is observed 
at the $\Gamma$ point
for Sn-doped CdSe and CdTe monolayers. Berry curvature values are more in all the Sn-doped monolayers 
than the pristine monolayers. These newly designed monolayers are thus
found to be suitable for valleytronics applications. Sn-doped monolayers show 
band inversion deep in the valence and conduction bands between Sn~$s$ and $p$ and X~$p$ states but
lack topological properties.  
\end{abstract}
\maketitle

\section{Introduction}
Two-dimensional (2D) structures form a class of materials which is very much in demand both from 
fundamental physics point of view and for applications.
Since the discovery of graphene monolayer in 
2004~\cite{novo}, 2D material family is increasing with fascinating electronic and 
spintronic properties. 2D Xene family members are extensively studied but are 
not stable in their free-standing form without substrates~\cite{chaip, vila, zhu}. 
Transition metal dichalcogenides (TMDs) are another group of 2D family with a
graphene-like hexagonal framework and have been realized experimentally~\cite{radi, zhang}. 
However, these TMDs are layered materials. The minimum possible thickness and wide band gaps of 2D 
semiconductors  make them important in 
the fields of photovoltaics, sensing, biomedicine, optoelectronics, photocatalysis, etc.

An interesting property seen in 2D materials is valleytronics in which
`valleys' are defined as the points in the energy-momentum plots having  local 
minima in conduction band (CB) or local maxima in valence band 
(VB)~\cite{rycerz}. This energy extremum or valley has an intrinsic magnetic moment associated with it
as the Bohr magneton is with electron spin, and offers a new degree of freedom of electrons. Similar to the 
properties like charge and spin, that are used in electronic and spintronic devices, this 
additional degree of freedom in 2D monolayers with special symmetry, the valley pseudospin, can be 
used in valleytronic devices. For 2D monolayers with hexagonal lattice, if the band edges are located 
at the corners of the Brillouin zone (+K  and -K  points), the existence of two or more degenerate 
but inequivalent valleys leads to valley physics. The two inequivalent valleys constitute a binary 
index making the use of valley index as a potential information carrier and has applications 
in information technology. Due to the large separation of +K  and -K  (or K$\textquotesingle$) 
valleys in the momentum space, the valley pseudospin is robust against low-energy phonons and 
impurity scatterings. Upon breaking space inversion symmetry, the carriers at the valleys are 
associated with the valley contrasting physical quantities. 
Valley physics properties are  extensively studied in MX$_2$ (M = Mo, W ; X = S, Se, Te) TMDs in 
their 2D form~\cite{Jiang, korm, Manzeli,Srivastava2015}. 
The inversion symmetry breaking and strong spin-orbit coupling 
(SOC) give rise to coupled spin and valley physics in hexagonal monolayers.
The spin-valley coupling at the valence-band edges suppresses spin
and valley relaxation independently, as flip of  spin or valley alone is energetically forbidden by the valley-contrasting spin splitting.
Valley Hall and spin Hall effects coexist in both
electron-doped and hole-doped systems.
These effects suggest the potential of integrated spintronic and valleytronic applications and 
the necessity and urgency to find and design 2D valleytronic materials with strong spin-valley coupling.

Cadmium chalcogenides CdX (X = S, Se, Te) in their bulk form are well-established wide band gap 
semiconductors with important applications in 
photovoltaics and optoelectronic and spintronic devices. 
Reducing the dimensions from 3D to 2D is expected to result in some exotic electronic and optical 
properties. 2D honeycomb-like free-standing single 
to few layered CdX sheets, studied theoretically, indicate that such materials are experimentally 
achievable~\cite{zhou, son}. However, 
there are no reports of synthesis of CdX monolayers experimentally till date. A possible reason is
the difficulty to use exfoliation technique 
since the layers are covalently bonded in the bulk structures. 
Few reports of synthesis of thick CdSe nanosheets and nanocrystalline Sn-doped CdSe films are 
available in the literature. Son \textit{et al.} have 
used soft template method to synthesize free-standing single layered lamellar-structured CdSe 
nanosheets with the diffraction pattern showing
growth along [000$\bar{1}$] and [1$\bar{1}$00] axes of hexagonal wurtzite structure~\cite{son}. 
Kaur \textit{et al.} have prepared
Sn-doped CdSe thin films on glass substrate by thermal evaporation which showed characteristic 
X-ray diffraction peaks of (100) plane of wurtzite
structure~\cite{kaur}. They have also concluded from X-ray diffraction studies that Sn$^{2+}$ 
ions substitute Cd$^{2+}$ ions in the CdSe lattice. On the
other hand, Sahu \textit{et al.} found tin-incorporated nanocrystalline CdSe thin films to be 
along cubic (111) plane~\cite{sahu}.  
Zinc blende (ZB) (111) plane and wurtzite (W) (001) plane have the same arrangement of atoms except that the 
stacking of layers is different.
Das \textit{et al.} have prepared thin films of Sn-doped CdS on glass substrate by controlled precipitation
of Cd$^{2+}$ and Sn$^{2+}$ ions simultaneously and have reported the analysis of the optical data based 
on Tauc relation~\cite{das}. They found that the energy band gap decreased from 2.1 to 1.9~eV and then to 1.85~eV 
as Sn composition in the solution is increased from 0\% to 1\% and then to 2\%.
Planar CdS monolayer is theoretically 
predicted and is found to be stable~\cite{pgarg}. 
Unsal \textit{et al.} have explored strain dependent properties of CdTe monolayer and they
found $\alpha$-PbO type planar structure of CdTe to be the most stable structure while planar
hexagonal CdTe to be dynamically unstable~\cite{unsal}.
Zheng \textit{et al.} have carried out a systematic study of 32 honeycomb monolayer II-VI semiconductors 
and their heterostructures for potential applications in optoelectronics, spintronics and strong 
correlated electronics~\cite{zheng}. However, they found 
the CdX sheets to be poor with respect to dynamic stability. Safari \textit{et al.} have employed full 
potential linear augmented plane wave method to study 
the bonding and optical transitions in planar ZnX and CdX monolayers~\cite{safari}. 

Mohanta \textit{et al.} have used density functional theory (DFT) coupled with semi-classical Boltzmann transport equations, to study
the transport properties of CdS and CdSe monolayers and have reported the sheets to be buckled for better 
stability~\cite{mohanta}. Ashwin Kishore \textit{et al.} have theoretically explored  CdX/C2N (X = S, Se) 
heterostructures as potential photocatalysts for water splitting~\cite{kishore}. Recently Opoku \textit{et al.}
have theoretically examined CdS/SiH heterostructure to study photocatalytic water splitting~\cite{opoku}.
Wang \textit{et al.} have examined the geometric, electronic, optical and potential photocatalytic properties of 
single-layer CdX  (X = S, Se, Te) sheets cleaved from the (001) plane of the bulk wurtzite structure~\cite{wang1}.
None of these studies explored the valley physics in these monolayers. 

Among all these earlier studies of CdX (X = S, Se, Te), in their 2D hexagonal form, only 
Zheng \textit{et al.} and Zhou \textit{et al.} have carried out comparative 
studies of CdS, CdSe and CdTe hexagonal monolayers but they did not find all the planar monolayers 
to be dynamically stable.
Our careful optimization and molecular dynamics studies indicate that all the pristine CdX hexagonal 
monolayers are almost planar and are structurally, 
dynamically and thermally stable. These are therefore good candidates to explore valley physics. 

SOC splits the bands into two spin sub-bands; this splitting is Zeeman-like since the bands are split as
if they are subjected to an external magnetic field. Zeeman-type splitting is a consequence of spin 
discrimination as a result of space inversion symmetry breaking at non-time reversal (non-TR) invariant 
$\vec{k}$-points and not due to an intrinsic electric dipole. Moreover, the valley also controls the 
spin-dependent properties and governs that the valence band maximum (VBM) and conduction band
minimum (CBM) must take place at a non-TR $\vec{k}$-point. The separation of the two spin-split bands
at the valleys is termed as valley spin-splitting (VSS). Rashba effect, on the other hand, is 
spontaneous spin-polarization coupled to the momentum of an electron emerging due to breaking of 
space inversion symmetry and SOC~\cite{rashba}. This spin degeneracy is lifted for the bands dispersing from the
center of the Brillouin zone (BZ) i.e., $\Gamma$ point or from the points at the BZ boundary, in the wave vector direction
and the spin is oriented in a direction perpendicular to both the surface normal and the momentum of the 
electron. The states exhibiting Rashba splitting consist of mainly p orbitals.

Recent works on suitable materials for spontaneous valley polarization~\cite{zhao}, anomalous valley Hall effect~\cite{peng}, strong spin-valley
coupling~\cite{dou}, large VSS~\cite{zang,ai} are worth mentioning and emphasize the need of finding suitable materials for valleytronics applications.

Can SOC initiate VSS in these CdX monolayers, 
if yes, how strong is VSS? are some of the questions we try to answer in this work. Since SOC strength 
is very important for significant VSS, we have further doped 
naturally abundant and nontoxic high Z material, tin, in  
these monolayers to enhance the effect of SOC. We do find that these Sn-doped CdX monolayers are also
structurally, dynamically and thermally stable.
We have then studied the electronic structure of all these monolayers to understand the role of SOC to
initiate VSS in these Sn-doped CdX monolayers. A comparative study of 
the electronic band structure of pristine and Sn-doped CdX monolayers brings out the role of Sn and shows 
that these Sn-doped CdX monolayers
show robust spin-valley coupling, larger VSS and Zeeman-like and Rashba splitting, especially in Sn-doped CdTe.
Our structural studies involving the cohesive 
energies reveal that Sn-doped CdX monolayers are more stable than their pristine counterparts. 
The formation energies indicate that Sn-doped CdX monolayers (in particular, Sn-doped CdTe) 
are viable in a bottom-up approach. These
observations may provide impetus for experimental realization of Sn-doped CdX monolayers.

\section{Computational Details}\label{method}
We have carried out the electronic structure calculations for the pristine and Sn-doped CdX monolayers  
within the framework of  density functional theory as implemented 
in Vienna Ab-Initio Simulation Package (VASP)~\cite{VASP2, VASP3}. The ion-electron interaction is described using 
projected augmented wave method~\cite{paw}. The exchange-correlation energies of the 
interacting electrons are considered using Perdew-Burke-Ernzerhof (PBE) generalized 
gradient approximation (GGA)~\cite{gga}. The Kohn-Sham wavefunctions are expanded in 
plane waves with kinetic energy cut-off  of 500~eV. The convergence threshold for self-consistent electronic 
energy calculations is set to 10$^{-6}$~eV 
in consecutive iterations. In order to 
project the DFT wave functions onto the maximally localized Wannier functions, 
we have used Wannier90 code~\cite{Most}.
PYPROCAR code is used for 
plotting the band structures~\cite{HERATH}. The dynamic stability
of the monolayers is established by calculating phonon dispersion  using the
density functional perturbation theory~\cite{baroni, gonze} as implemented
in the VASP code along with the PHONOPY package~\cite{togo}.   We have also 
investigated the thermal stability with ab initio molecular dynamics 
(AIMD)~\cite{ang} using Nos\'{e} 
thermostat~\cite{nose} for NVT
ensemble at 300~K. The simulation is carried out for 45~ps with a time step of 3.0~fs 
Bader charge analysis is carried out to calculate the charge transfer amongst the atoms~\cite{Tang,bader}.
The Bader charge for an atom is the difference between the number of its valence electrons and its total 
charge calculated from the Bader 
charge analysis. Ionicity of bonds is decided by nonzero Bader charges.  

\begin{figure*}
\vskip -0.471in
\hskip 0.2in
\includegraphics[scale=0.16]{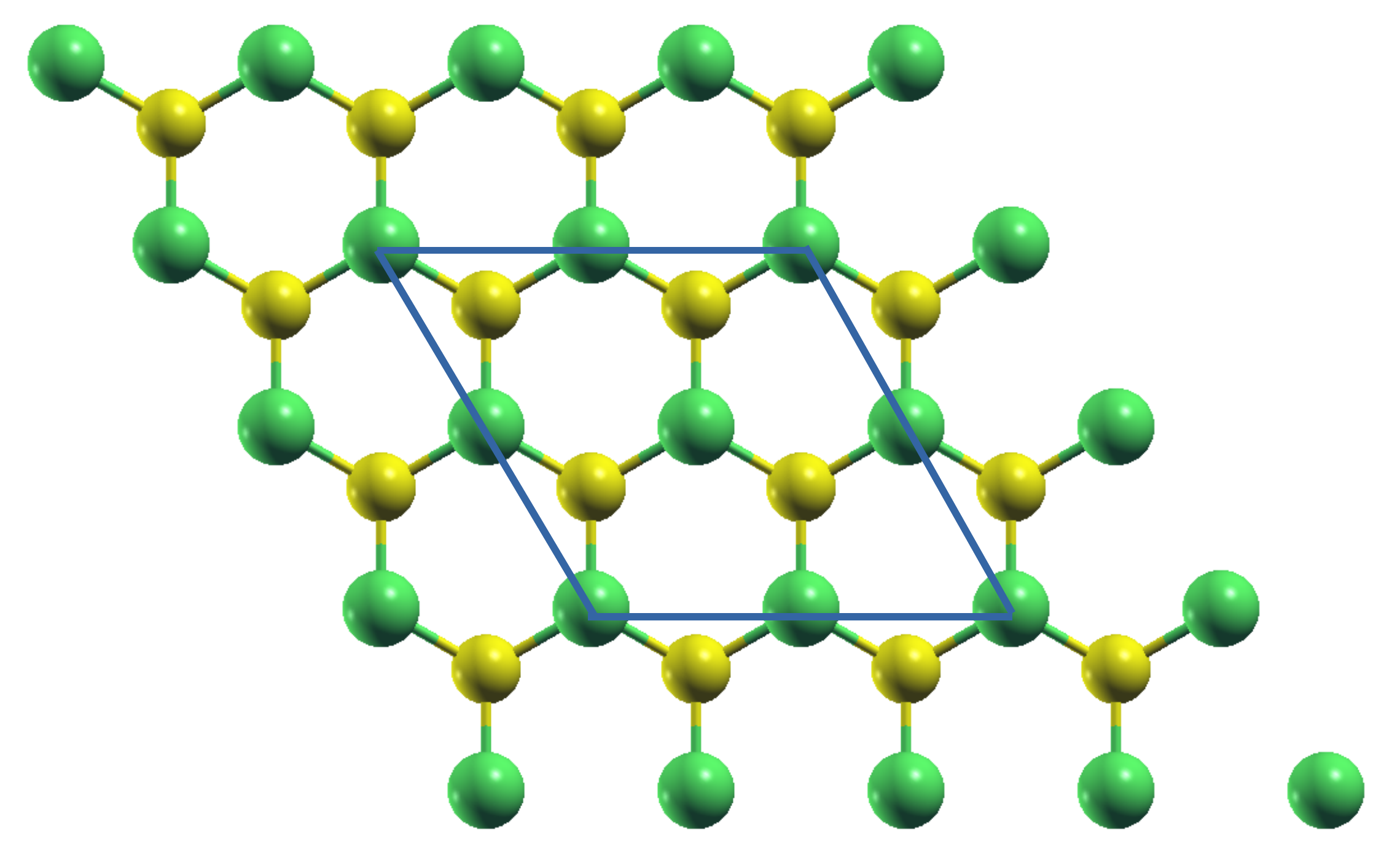}  \hskip 0.35in
\includegraphics[scale=0.15]{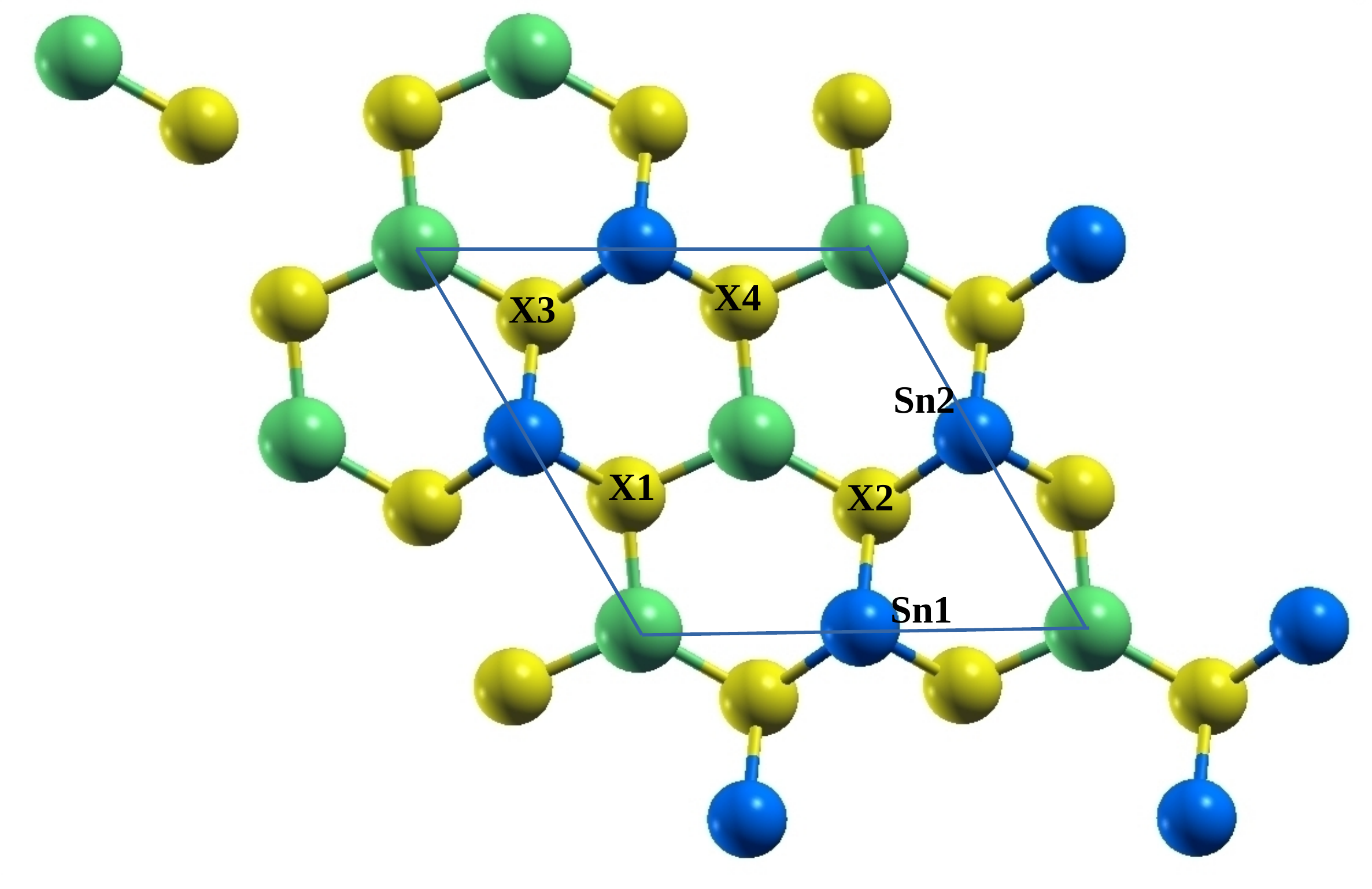}\\
.\hskip 0.5in (a) \hskip 2.9in (b)\\
\vskip 0.12in
\hskip 0.4in
\includegraphics[scale=0.14]{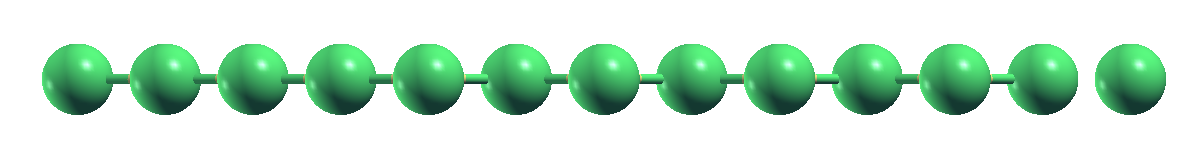}  \hskip 0.6in
\includegraphics[scale=0.14]{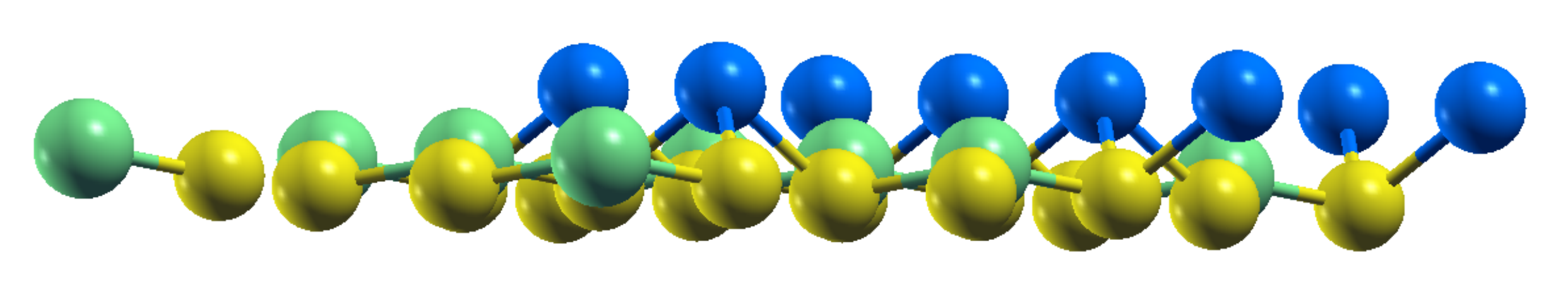}\\
.\hskip 0.5in (c) \hskip 2.9in (d)\\
\hskip -0.04in
\includegraphics[scale=0.12]{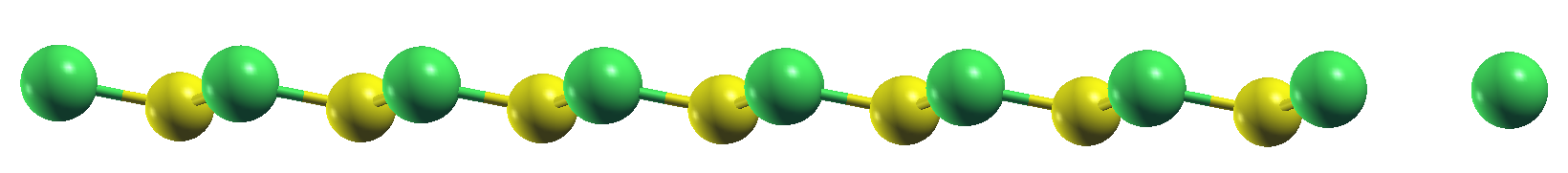} \hskip 0.5in\includegraphics[scale=0.14]{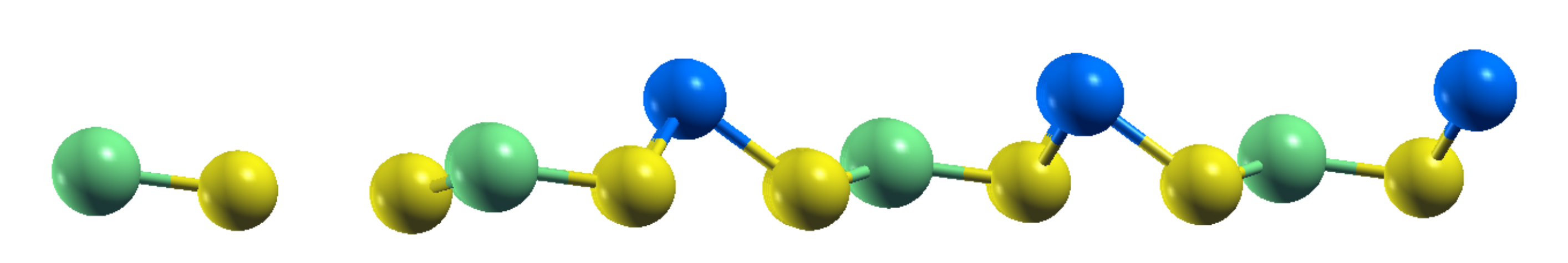}\\
.\hskip 0.5in (e) \hskip 2.9in (f)\\
\vskip 0.1in
\includegraphics[scale=0.3]{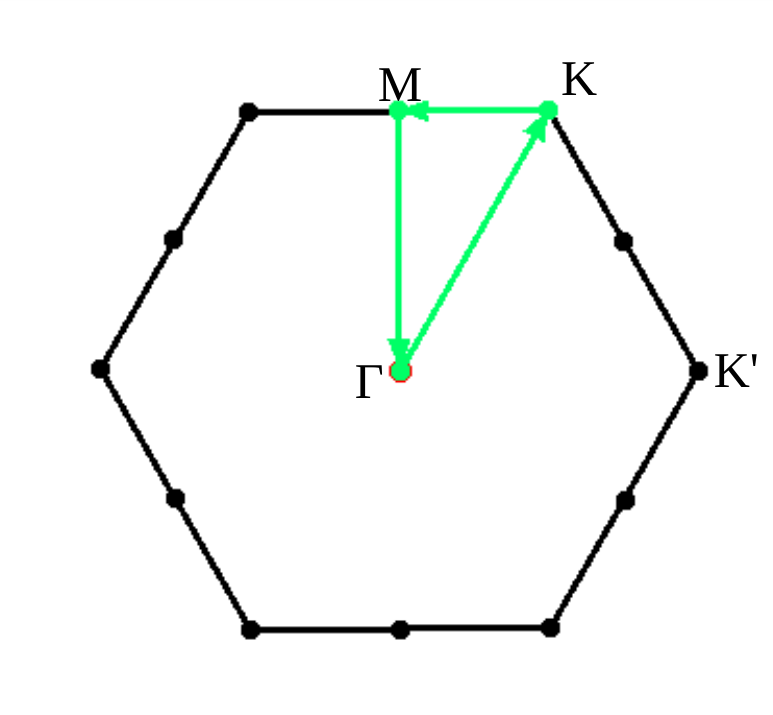}\\
\vskip -0.1in
 \hskip 0.2in  (g)
\caption{(Color online) Geometry  of structurally relaxed  pristine and Sn-doped CdX monolayers. 
Green, yellow and blue colored balls respectively represent cadmium atoms, chalcogen atoms 
and tin atoms. (a) and (b) represent the top views in $xy$-plane of pristine and Sn-doped 
monolayers while (c) and (d) represent 
the respective side views with Sn atoms out of the $xy$-plane showing buckling along 
$z$-axis and (e) and (f) are corresponding tilted views for better visualization. 
The unit cells are shown in (a) and (b) by blue lines. The Brillouin zone is depicted in
(g) with labels of the high symmetry points.} 
\label{geom}
\end{figure*}

In this work, we have constructed free-standing pristine CdX monolayers from 
(111) surface of CdX zinc blende (ZB) structure in the $xy$-plane as shown in Fig.~\ref{geom}(a).  We
have then doped Sn in CdX monolayers by substitution at Cd site as shown in Fig.~\ref{geom}(b).
It may be noted that the experimental study by Sahu \textit{et al.} of tin-incorporated nanocrystalline 
CdSe thin films have reported the structure to be (111) plane of cubic ZB structure~\cite{sahu}
and Kaur \textit{et al.} have observed that Sn$^{2+}$ ions substitute Cd$^{2+}$ in the 
CdSe lattice~\cite{kaur}. A $2\times2$ planar unit cell   
is shown in Fig.~\ref{geom}(a) and Fig.~\ref{geom}(b) for the pristine and Sn-doped monolayers respectively. 
In order to avoid interaction of
the  monolayer with its periodic images, we have kept vacuum of minimum 15~\AA\ 
in the $z$-direction. The self-consistent calculations are performed with 
$8\times8\times1$ mesh of $\vec{k}$-points using $\Gamma$-centered Monkhorst-Pack scheme. 
All pristine and Sn-doped CdX monolayer structures are fully relaxed using the conjugate gradient method,
keeping only the volume of the unit cell constant. All the atoms in the monolayer are allowed to move
in any direction and the structure is considered to be optimized when the force 
on each atom is less than 0.001~eV/\AA. After relaxation, a $2 \times 2 \times 1 $ supercell of all the  
monolayers is used for  the calculation of phonon dispersion spectra.

\section{Results and Discussions}
CdX monolayer can be considered to be isovalent with other graphene analogues 
(leaving the 4$d^{10}$ shell of Cd) and therefore honeycomb lattice is a natural choice~\cite{zhou}.
The atomic structure of fully relaxed pristine CdX  monolayers is almost planar 
hexagonal as shown in Fig.~\ref{geom}(a), (c) and (e). The lattice parameters obtained after 
structure optimization are  in good agreement with previous 
calculations~\cite{pgarg, opoku, mohanta, kishore, safari, zheng, zhou, wang1}. The monolayers
exhibit space group P3m1. 
Fully relaxed structures of Sn-doped CdX monolayers 
show slight buckling with the Sn atoms coming out of the $xy$-plane as shown 
in Fig.~\ref{geom}(d) and (f). Due to the larger size of the Sn atom in comparison
to Cd atom, the buckled structure exhibits a stable structure.
The magnitudes of buckling index $d_z$, along with the lattice constants for pristine and Sn-doped 
CdX monolayers  are listed in Tables S1 and S2 in supplementary information (SI) along with the 
relevant data available from previous published works on pristine systems.  Due to buckling in the 
$z$-direction, Sn-doped CdX monolayers show reduced hexagonal symmetry with the space group P6m and 
the lattice parameters in the $xy$-plane are reduced in comparison to the respective pristine structures.
Such decrease in lattice parameters is reported by Sahu \textit{et al.} for tin-incorporated CdSe 
nanocrystalline thin films~\cite{sahu}. Both pristine and Sn-doped monolayers lack inversion symmetry. 
2D Brillouin zone, with three high symmetry edge $\vec{k}$-points, M, K and K$\textquotesingle$, 
is shown in Fig.~\ref{geom}(g).

The relaxed structures of all the pristine and Sn-doped monolayers are checked for dynamic 
stability by calculating the  phonon band structure (shown in Fig.~S1 in SI) and 
the results do not show signature of any imaginary mode. Zheng \textit{et al.}~\cite{zheng} and Unsal 
{\textit{et al.}}~\cite{unsal} have studied planar structures of CdX (X = O, S, Se, Te) and CdTe respectively but
found that these structures are not dynamically stable. However, we found our almost planar structures 
to be dynamically stable. 
There are total 24 branches with 3 acoustic
phonon modes and 21 optical phonon modes. The highest phonon frequency decreases as the
atomic weight of X atoms increases from S to Te in CdX and Sn-doped CdX monolayers. 
All the monolayers, studied in this work, satisfy the Born-Huang criteria :
$C_{11}C_{22} - {C_{12}}^2 > 0$ and $C_{66} > 0$, which indicates the mechanical stability of the monolayers.

Cohesive energy (per atom), as defined in Eqs.~(\ref{coh1}) and (\ref{coh2}), 
of a monolayer is a measure of binding between the atoms 
constituting the monolayer and indicates the possibility of monolayer 
formation if the values are positive.
The values of cohesive energy $E_{coh}^{CdX}$  for pristine
and $E_{coh}^{Sn-doped CdX}$ for Sn-doped monolayers are tabulated in Table~\ref{form} and are indeed found to be positive.
Previous results are also quoted for comparison.
\begin{equation}
 E_{coh}^{CdX}=\frac{n*({E_{Cd}} +E_X)-E_{CdX}^{mono}}{2n}
\label{coh1}
\end{equation}
and
\begin{eqnarray}
  & E_{coh}^{Sn-dopedCdX}=~~~~~~~~~~~~~~~~~~~~~~~~~~\\ [10pt] \nonumber
  & \frac{(n-m)*{E_{Cd}} +n*E_X +m*E_{Sn}-E_{Sn-dopedCdX}^{mono}}{2n}
\label{coh2}
\end{eqnarray}
\noindent
where $E_{Y}$ is the total energy of the isolated  atom Y, $E_{CdX}^{mono}$
is the total energy of monolayer consisting of $n$ atoms each of Cd and X 
and $E_{Sn-dopedCdX}^{mono}$ is the total energy of above CdX monolayer in which  $m$ atoms of 
Cd are replaced by Sn atoms. 

\begin{table*}
\caption{Cohesive energy per atom, $E_{coh}^{CdX}$ and $E_{coh}^{Sn-dopedCdX}$ and formation energy 
per atom (from constituents), $E_{for}^{CdX}$ and $E_{for}^{Sn-dopedCdX}$, of pristine CdX and Sn-doped CdX
monolayers are listed in eV. Formation energies per atom (from bulk) $\Delta E_{coh}^{CdX}$ of the pristine
CdX monolayers are also given in eV. Cohesive energies 
per atom, $E_{cohbulk}^{CdX}$, and formation energies per atom (from constituents),
$ E_{forbulk}^{CdX}$,
for bulk CdS, CdSe and CdTe are 2.90~eV, 2.56~eV, 2.23~eV and 0.636~eV, 0.617~eV, 0.457~eV respectively 
from our calculations.
Values of relevant quantities available from literature are also included for comparison.}

\begin{center}
\begin{tabular}{|c|c|c|c|c|c|c|}
\hline
 &   \multicolumn{3}{|c|}{Pristine CdX monolayer} &  \multicolumn{3}{|c|}{Sn-doped CdX monolayer}\\
    \hline
&~CdS~&~CdSe~&~CdTe~&~CdS~&~CdSe~&~CdTe~\\
\hline
 $E_{coh}^{CdX}$ or $E_{coh}^{Sn-dopedCdX}$ & 2.66 & 2.33 & 2.00 & 3.29 & 2.98 & 2.62 \\
 \hline
 $E_{coh}^{CdX}$ from literature&2.65$^a$,2.93$^c$&2.66$^c$&2.37$^c$&-&-&-\\
 & 3.5$^d$ & 2.99$^d$ & 1.90$^b$, 2.45$^d$& & &\\
 \hline 
$E_{for}^{CdX}$ or $E_{for}^{Sn-dopedCdX}$ &0.403  &0.385  & 0.230 & 0.335 & 0.339  & 0.158 \\
\hline
$\Delta E_{coh}^{CdX}$ &0.233 &0.231&0.228 &- &- &-\\
\hline
$\Delta E_{coh}^{CdX}$ from literature&0.38$^c$, 0.20$^e$,&0.385$^c$,&0.366$^c$,&-&-&-\\
&0.54$^d$ & 0.21$^e$& 0.18$^e$&&&\\
\hline
\end{tabular}
\end{center}
\label{form}
a: Ref.~\cite{pgarg}, b: Ref.~\cite{unsal}, c: Ref.~\cite{zheng}, d: Ref.~\cite{safari}, e: Ref.~\cite{wang1}.
\vspace{-0.2in}
\end{table*}

Likewise, cohesive energy of bulk CdX is defined by Eq. ({\ref{coh3}) 
\begin{equation}
 E_{cohbulk}^{CdX}=\frac{n*({E_{Cd}} +E_X)-E_{CdX}^{bulk}}{2n}
\label{coh3}
\end{equation}
 where $E_{CdX}^{bulk}$ is the total energy of bulk CdX consisting of $n$ atoms each
 of Cd and X.

Cohesive energy of bulk CdX decreases from S to Te and the same trend is seen in the cohesive energy values for
pristine and Sn-doped monolayers. It may be noted that cohesive energy of each Sn-doped
monolayer is more than its pristine counterpart by almost $\sim$~0.6~eV. To understand the reason why 
it is so, we have compared the difference charge densities (charge density of monolayer - sum of the charge densities of atoms constituting the monolayer)(please refer to SI) of pristine and
Sn-doped CdX monolayers. This comparison brings out the changes in bonding on Sn-doping.
It is seen that (i)~there is more charge build-up between X and Sn atoms in comparison to X and Cd 
atoms and  (ii)~there is charge depletion between the two Sn atoms which is absent in pristine 
structures between two Cd atoms. This results in stronger binding between the 
atoms in Sn-doped monolayers and hence these monolayers are  more stable.

\begin{equation}
\Delta E_{coh}^{CdX}=E_{cohbulk}^{CdX}-E_{coh}^{CdX}
\label{delform}
\end{equation}
\noindent
$\Delta E_{coh}^{CdX}$, the difference between cohesive energies of bulk and monolayer, 
as defined in Eq.~(\ref{delform}), is the
formation energy per atom (from bulk) and is the cost to be 
borne to form a 2D structure from the bulk structure~\cite{zheng}.
Obviously, cohesive energies of monolayers are smaller than their corresponding bulk
structures since the monolayers do not exist naturally. Our  values of 
$\Delta E_{coh}^{CdX}$, as tabulated in Table~\ref{form}, are less than two-third the 
values quoted by Zheng~\textit{et al.}~\cite{zheng} indicating that our 2D structures
are more stable than those predicted by Zheng~\textit{et al.}. 
They have obtained CdS monolayer to be planar but CdSe and CdTe to be buckled
structures. It may be noted here that 
Zheng~\textit{et al.}~\cite{zheng} found their 2D CdX structures to be poor with respect to dynamic stability.
Dynamic stability of our structures is confirmed and  indicates that the 2D 
structures designed in this work lie at least at a local minimum of potential energy 
surface rather than at a saddle point~\cite{zhuang}.
 It may be mentioned that formation energy per atom (from bulk) of single
layer of ZnO, which has been successfully synthesized~\cite{tusche}, has been calculated 
to be 0.355~eV by Zheng~{\textit{et al.}}~\cite{zheng}. 
Our values of $\Delta E_{coh}^{CdX}$ are smaller than this value indicating the feasibility of 
experimental realization of CdX monolayers.

A monolayer, when prepared from its elemental atoms, requires some energy cost, 
which is quantified as the formation energy by Daguer~\textit{et al.}~\cite{daguer}. The formation energies 
(per atom) (from constituents) of the pristine and Sn-doped monolayers are defined as 
\begin{equation}
 E_{for}^{CdX}=\frac{2n*E_{coh}^{CdX}-[n*(E_{cohbulk}^{Cd} + E_{cohbulk}^{X})]}{2n}
 \end{equation}
    and
  \begin{eqnarray}
 &E_{for}^{Sn-dopedCdX} = ~~~~~~~~~~~~~~~~~~~~~~~~~~~~~~~~~~~~~~~~~~~~  \\[10pt] \nonumber
 &\frac{2n*E_{coh}^{Sn-dopedCdX}-
[(n-m)*E_{cohbulk}^{Cd}+n*E_{cohbulk}^{X}+m*E_{cohbulk}^{Sn}]}{2n} 
\end{eqnarray}
\noindent 
where $ E_{cohbulk}^Y= E_Y -E_{Y}^{bulk} $ and $E_{Y}^{bulk}$ is the total energy per atom
of elemental bulk system Y.

The formation energies for all the monolayers, as tabulated in Table~\ref{form},
are positive indicating that monolayers do not form naturally. 
However, the monolayer structures have lower formation energy than their bulk structures 
for pristine materials suggesting their syntheses are feasible. The Sn-doped monolayers 
have lower formation energy than the corresponding pristine monolayers indicating 
Sn-doped monolayers can be achieved experimentally. The values of cohesive 
energies and formation energies, obtained in the present work, are compared
with earlier published results whenever available~\cite{pgarg, unsal, zheng, safari, wang1} and 
are in fair agreement as seen in Table~\ref{form}.
Comparing the values of formation energies per atom (from bulk)$\Delta E_{coh}^{CdX}$ and (from constituents) $E_{for}^{CdX}$
for the pristine structures, it is found that the latter is smaller for CdTe monolayer while the 
values are almost the same for CdSe monolayer
and $E_{for}^{CdX}$ is larger for CdS.
Thus, it seems that CdTe monolayer is easy to form from the constituents while CdS monolayer is easier to form from the bulk.
Also, the values of $E_{for}^{Sn-dopedCdX}$ are smaller
than the corresponding values of $E_{for}^{CdX}$,
indicating that it may be easier to synthesize the Sn-doped monolayers from the constituents using methods
like molecular beam epitaxy, 
chemical vapour deposition, metallorganic chemical vapor deposition etc. 

To investigate the thermal stability of Sn-doped monolayers, molecular dynamics 
simulations are performed at 300K and the variations in total energy and temperature as a 
function of time step are shown in Fig.~S3 of  SI along with the snapshots of the resulting 
final structures. The plots indicate that the monolayers are thermally stable.

\begin{figure*}
\vspace{-0.2in}
\hspace*{-0.25in}
\includegraphics[scale=0.22]{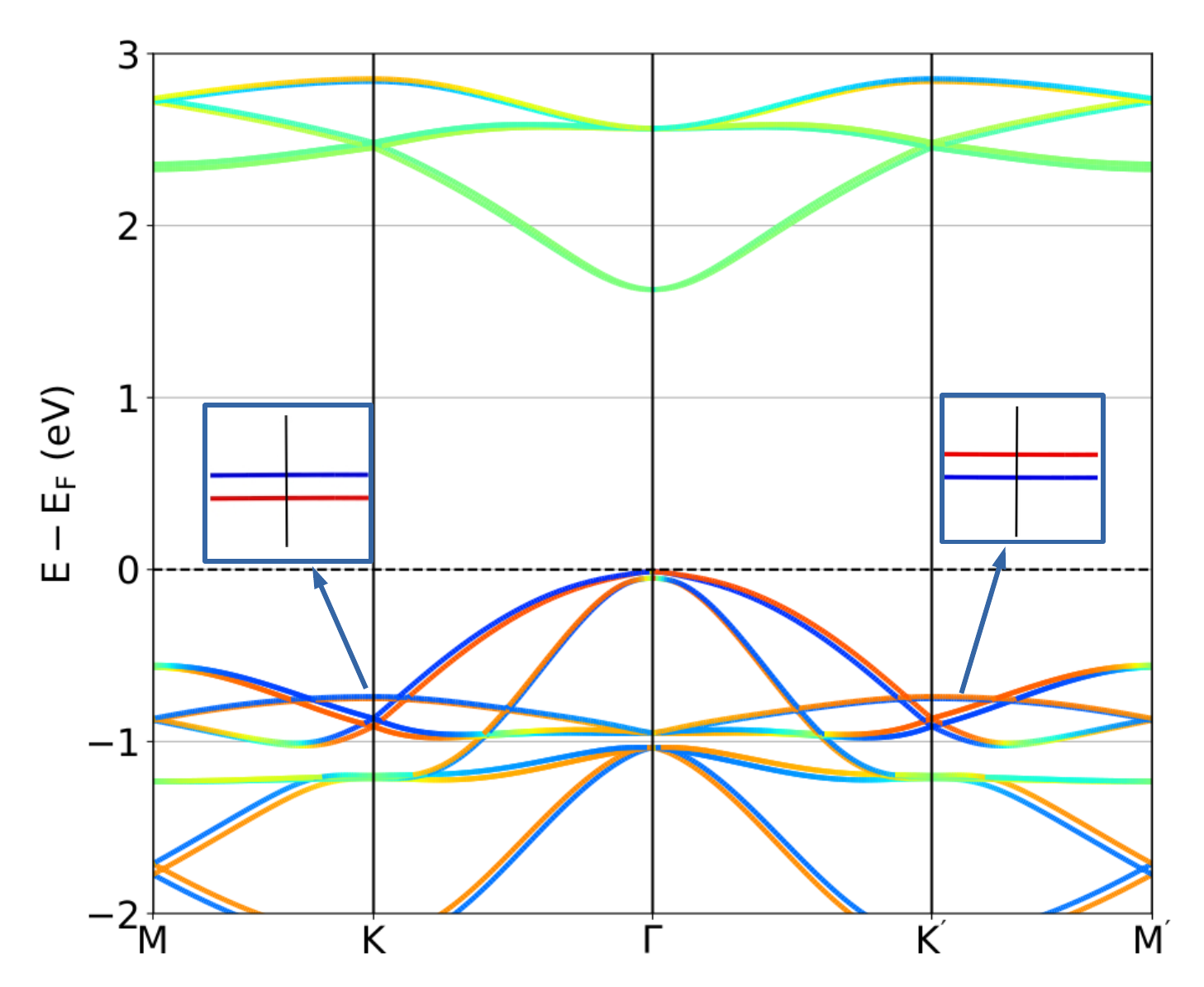} \includegraphics[scale=0.22]{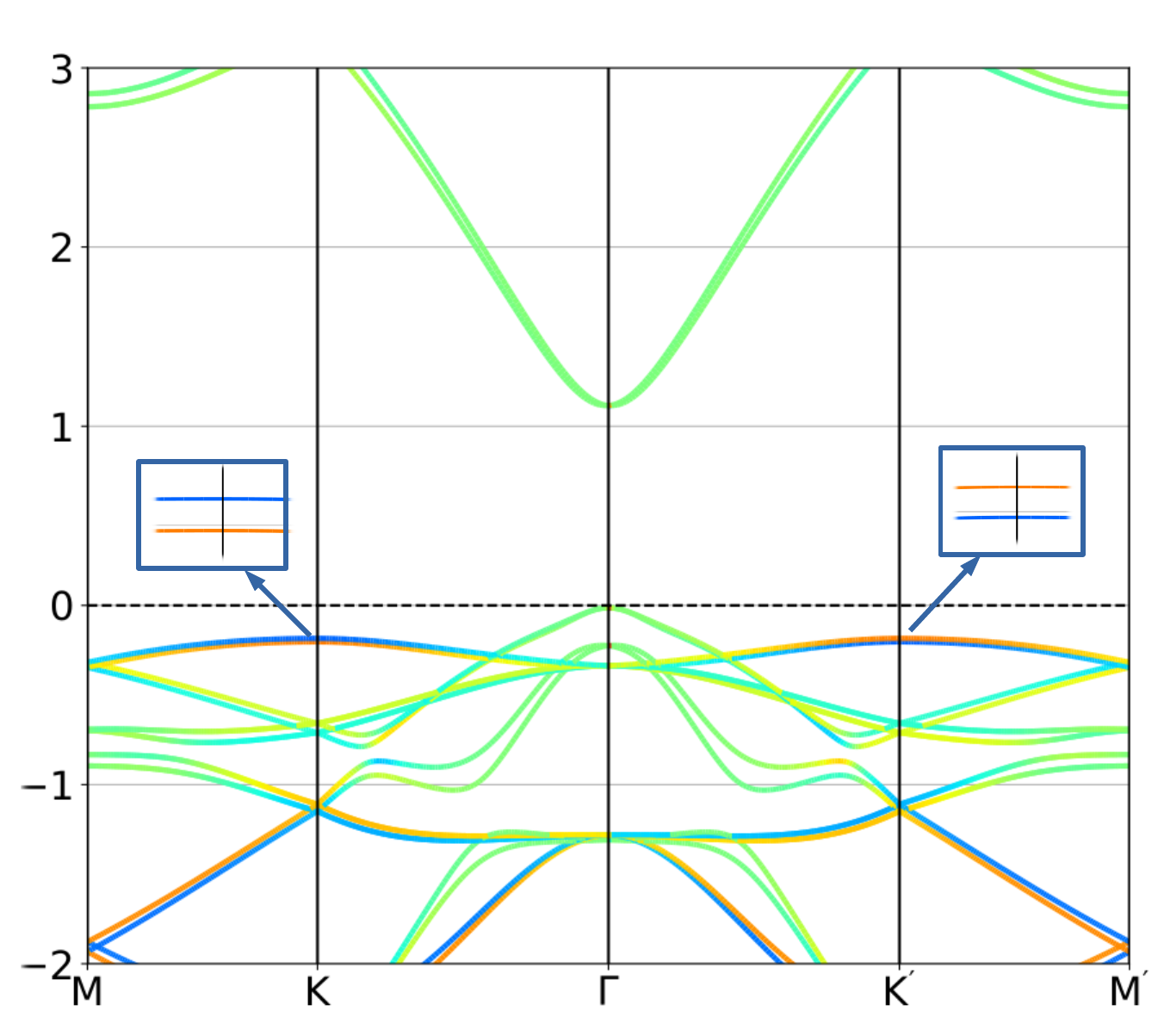}
\includegraphics[scale=0.22]{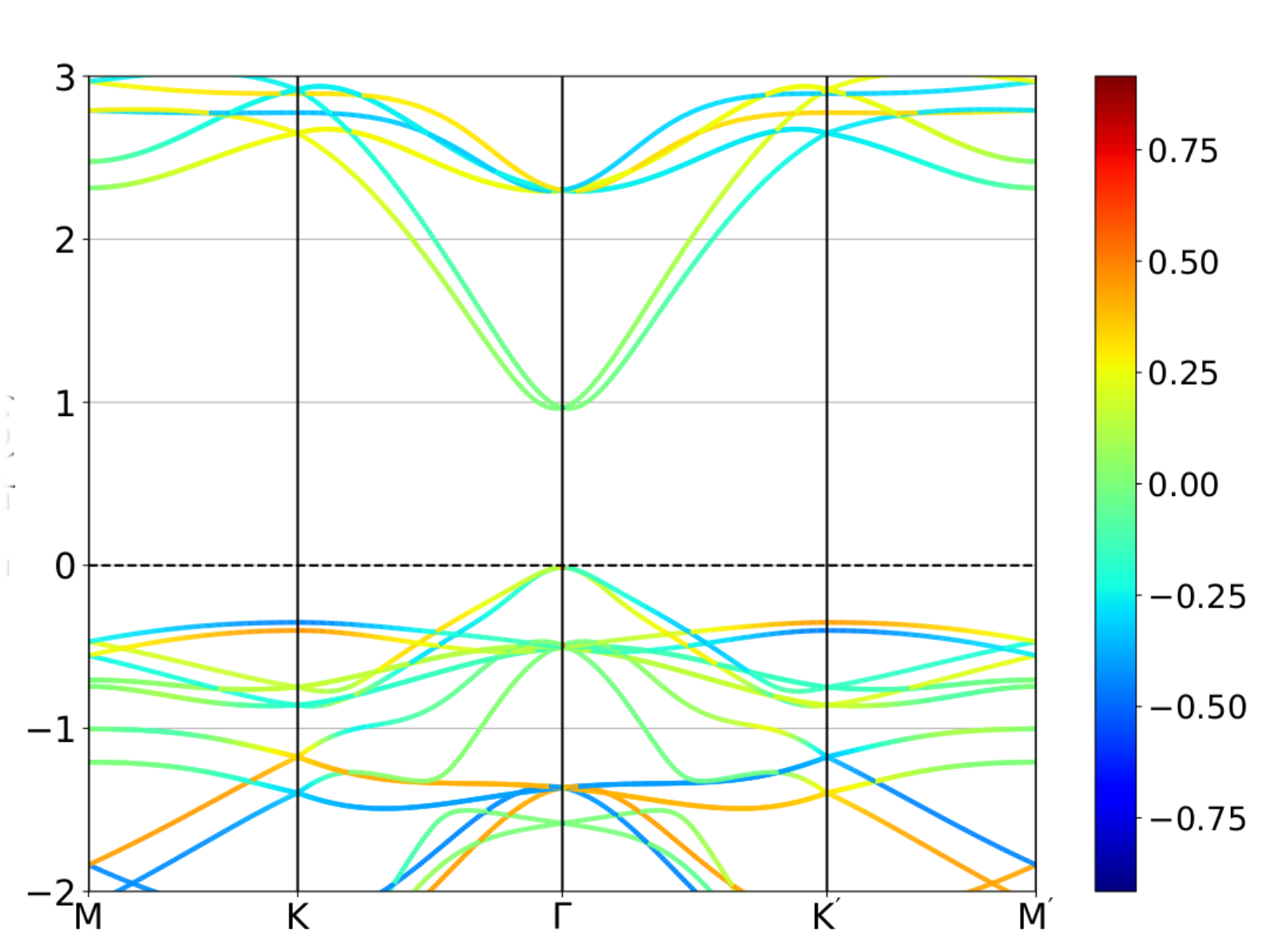}\\
  \hskip -0.4in  (a) \hskip 1.8in (b) \hskip 1.8in (c) \\  
 \vspace{-0.15in}
\caption{(Color online) Band structure plots, with inclusion of SOC, depicting spin-splitting in pristine  
CdX (X = S, Se and Te) monolayers  
are shown in (a), (b) and (c) respectively.  The respective Fermi energy is set to zero 
in all plots. Insets in (a) and (b)
show the enlarged view  indicating spin-splitting and spin-flip at K and K$\textquotesingle$ 
valleys. This is clearly visible in (c) for CdTe. Colors quantify
the expectation value of the $S_z$ component of spin for the bands. The extreme values with red and blue colors 
indicate up and down directions respectively.}
 \label{band-prestine-soc} 
 \vspace{-0.15in}
\end{figure*}

Pristine CdS and CdSe monolayers are direct band gap semiconductors for calculations without and
with inclusion of  SOC, however CdTe turns into an indirect band gap semiconductor with inclusion of 
SOC as shown in Fig.~\ref{band-prestine-soc} (and Fig.~S4 in SI). Band gap decreases from CdS to CdTe 
due to more extended $p$ orbitals of the 
chalcogen atoms in moving from S to Te. The lowest conduction band, formed mainly by the
Cd $s$ orbitals (hybridized with the chalcogen $s$ and $p$ orbitals) is also more dispersive.   
Our calculated band gaps without SOC for pristine monolayers are in good agreement 
with previous calculations~\cite{zhou, pgarg, zheng, safari, mohanta, kishore, wang1} (Please see Table S1 in SI).
One Cd atom has  
three X near-neighbor atoms forming trigonal planar structure. The $p$ orbitals of Cd and 
X atoms split into $p_x, p_y$ and $p_z$ components under the crystal field. The valence band (VB) region 
near the Fermi energy is mainly
formed by the X $p$ orbitals (the top valence band involves $p_z$ orbitals) hybridized with Cd $s$ and $p$ orbitals 
(very small contribution) as revealed from the projected density of states (PDOS) plots depicted in  
Fig.~S5 in SI. The lowest conduction band (CB) is formed
by Cd $s$ with little contribution of X $s$ and $p$ orbitals. This is a result of confinement in 2D structures. 
In bulk CdX, the valence band maximum (VBM) originates purely from X $p$ states and
conduction band minimum (CBM) is formed by Cd $s$ states.  
The hybridization near VBM is $sp^3$ in bulk CdX while it changes to $sp^2$ for monolayers. This transition from
$sp^3$ to $sp^2$ results in shortening of Cd-X distance to compensate for the reduced coordination. 
The CBM has antibonding $s$ states of Cd.

All pristine 
structures do not show band inversion even after inclusion of spin-orbit coupling. 

Band degeneracy is lifted with the application of SOC at K and K$\textquotesingle$ symmetry points
as shown in Fig.~\ref{band-prestine-soc}(a) to (c).  The band 
splitting is very prominent in CdTe monolayer in comparison to CdSe and CdS monolayers.
High atomic number of Te  results in strong SOC effect, 
causing prominent band splitting as evident from the band structure plot in 
Fig.~\ref{band-prestine-soc}(c). Spin-flip at K and K$\textquotesingle$ is clearly visible for CdTe and 
in the insets for CdS and CdSe. This spin-splitting at the valleys with inclusion of SOC in the calculations
is called valley spin splitting (VSS).
All the pristine monolayers show nice valley-spin coupling at K and K$\textquotesingle$ points but the 
VBM does not occur at K and K$\textquotesingle$ points. In order to confirm the 
spin component at the high symmetry points, we have 
calculated the expectation values of the spin operators ($S_x$, $S_y$ and $S_z$).
The bands at the K and K$\textquotesingle$ points arise due to X $p_z$ states and
the corresponding $S_z$ component is showing
purely $+ve$ and $-ve$ orientations at K and K$\textquotesingle$ valleys
respectively as revealed in Table S4 in SI. With inclusion of SOC and in presence of inversion asymmetry, 
the valley degeneracy at K and 
K$\textquotesingle$ points is lifted and the bands are spin-split. This splitting  is similar to one  
observed in magnetic Zeeman
effect but occurs in absence of magnetic field. This type of spin-splitting without application of 
magnetic field is called as 
Zeeman-type spin-splitting which lifts the out-of-plane spin degeneracy of the top most valence bands at K and 
K$\textquotesingle$ points. 
Due to the presence of time-reversal symmetry, the spin ordering on splitting is opposite in valleys 
at K point and its 
time reversed image point K$\textquotesingle$, as depicted by the colors of the  bands in Fig.~\ref{band-prestine-soc}. 
This shows a strong spin-valley coupling. 

\begin{figure*}
\hspace*{-0.25in}
\includegraphics[scale=0.22]{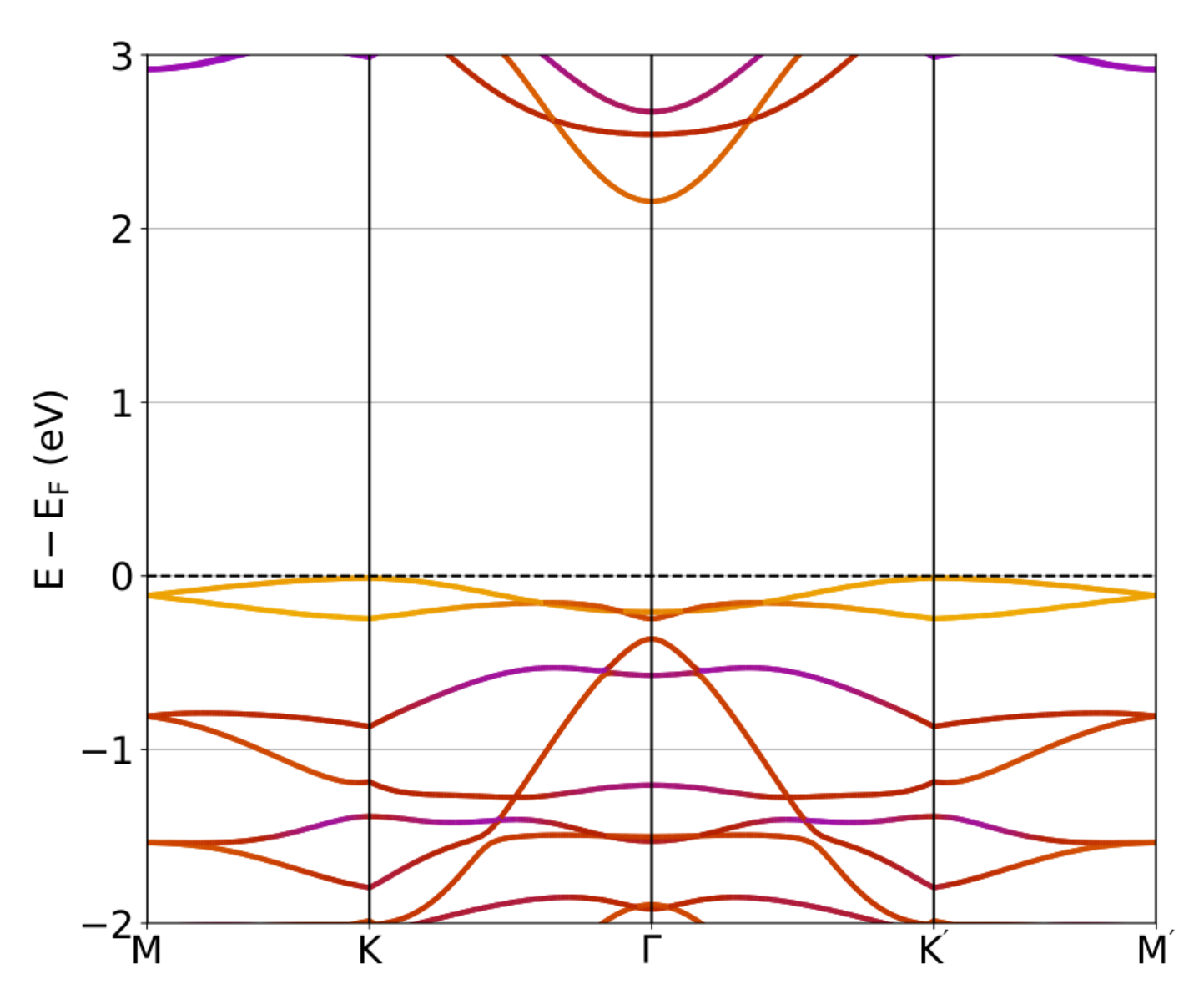} \includegraphics[scale=0.22]{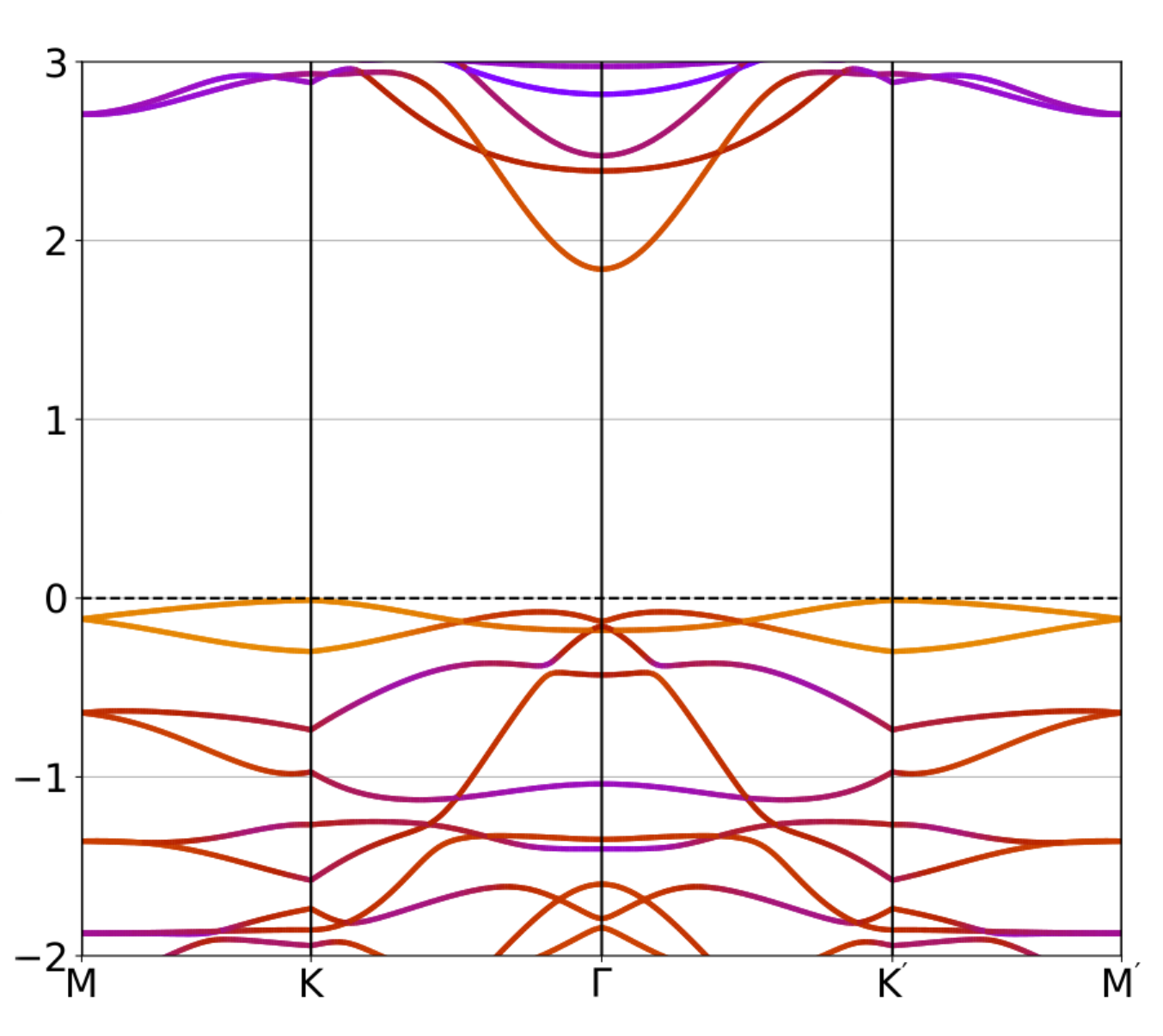}
\includegraphics[scale=0.22]{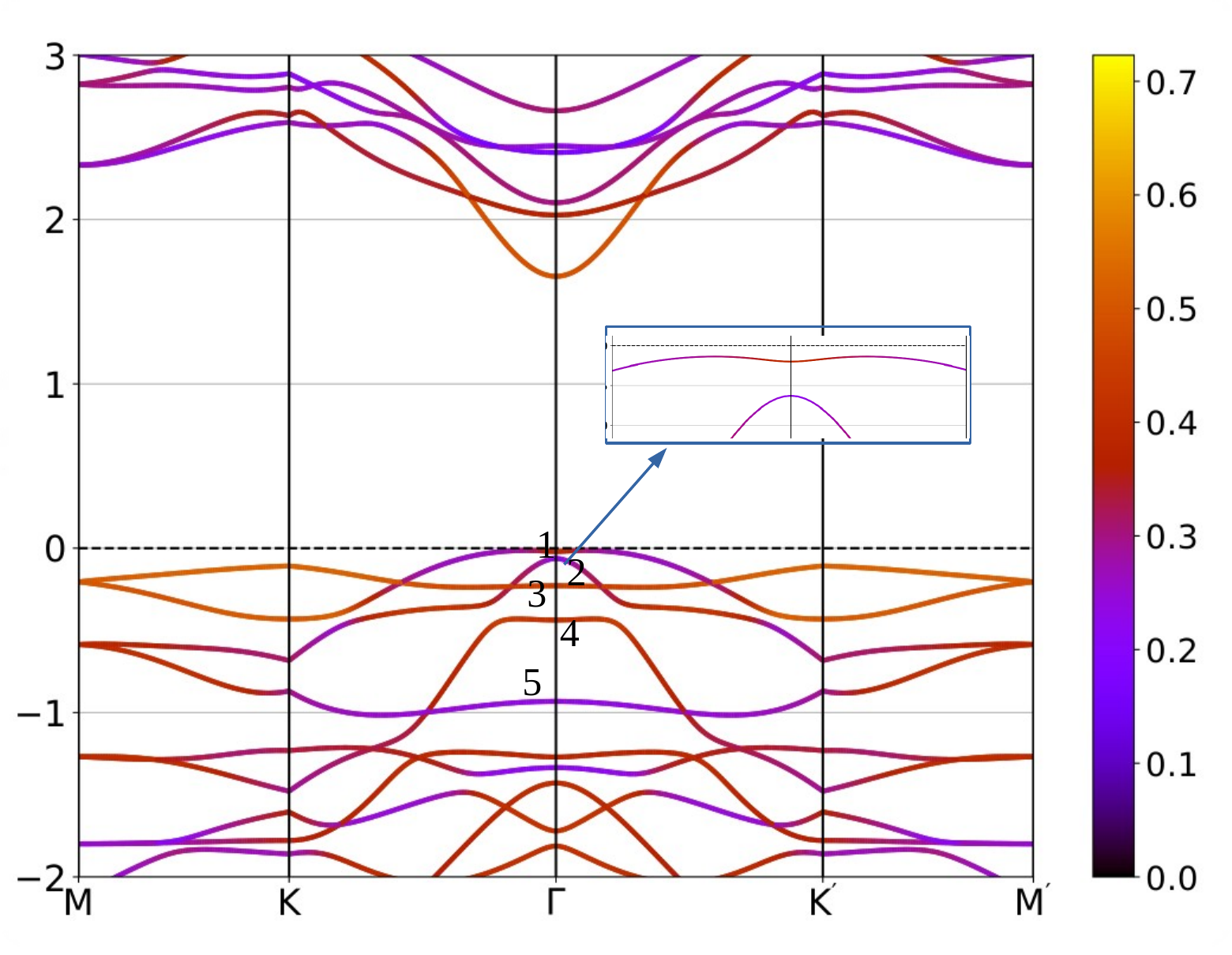} \\
 \hskip -0.4in  (a) \hskip 1.8in (b) \hskip 1.8in (c) \\
 \vspace{-0.15in}
\caption{(Color online) Band structure plots (without inclusion of SOC) for Sn-doped CdX (X = S, Se and Te) monolayers 
are shown in (a), (b) and (c) respectively. The respective Fermi energy is set to zero in all plots. 
The colormap goes from 0.0 (maximum $s$ orbital projection) to 0.7 (maximum $p$ orbital projection). 
Band inversion occurs at and near $\Gamma$  point between S $p$ and Sn $p$ states in (a) while
between Cd $s$, X $p$ and Sn $s$ and $p$ states in (b) and (c). The inset in (c) depicts the expanded
view of the band inversions. The 
information about the atomic states is obtained from the PDOS plots (Fig.~S7 in SI).}
\label{monolayer-bands}
\vspace{-0.15in}
\end{figure*}

The electronic band structures for Sn-doped CdX monolayers without inclusion of SOC (Fig.~\ref{monolayer-bands}) 
reveal that all monolayers are indirect band gap semiconductors  with increased band gap values
than their pristine counterparts as given in Table S2 in SI. 
This observation is in contradiction with the experimental band gaps reported in literature~\cite{kaur, sahu, das}. 
However, the experiments are not for monolayers and have only 2-5~\% Sn-doping.
The band gaps with inclusion of SOC are slightly less than those without inclusion of SOC. 
All Sn-doped monolayer systems show band inversion character in the valence band region around 
$\Gamma$ point (as seen from the different colors of bands). 
A specific band inversion  is shown in Fig.~\ref{monolayer-bands}(c) in an enlarged view. 
This band inversion between two bands 1 and 2 at $\Gamma$ point is due to $s$ and $p_z$ states of  
Sn and X atoms, while the band inversion between the two bands 4 and 5 at $\Gamma$ point is due to 
$p_z$ states of atoms X2 and X3 and $s$ states of Sn atoms. (Please refer Fig.~\ref{geom}(b) for the labels of atoms). 
The PDOS plots shown in Fig.~S7 in SI reveal strong hybridization of Sn $s$ and $p$ states with 
X $p$ states which is responsible for band inversion in all the Sn-doped monolayers.

\begin{figure*}
\hspace*{-0.25in}
\includegraphics[scale=0.22]{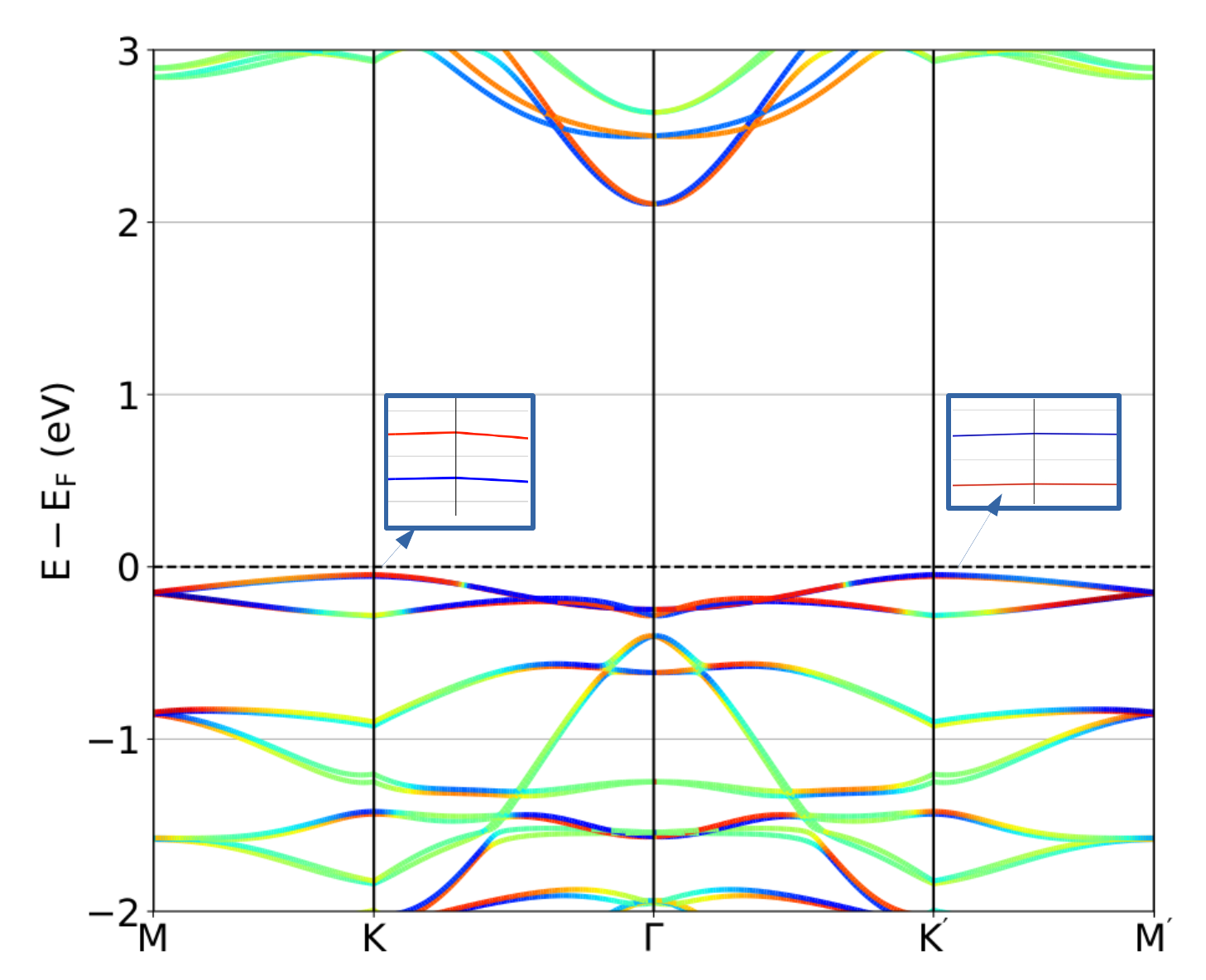}
\includegraphics[scale=0.22]{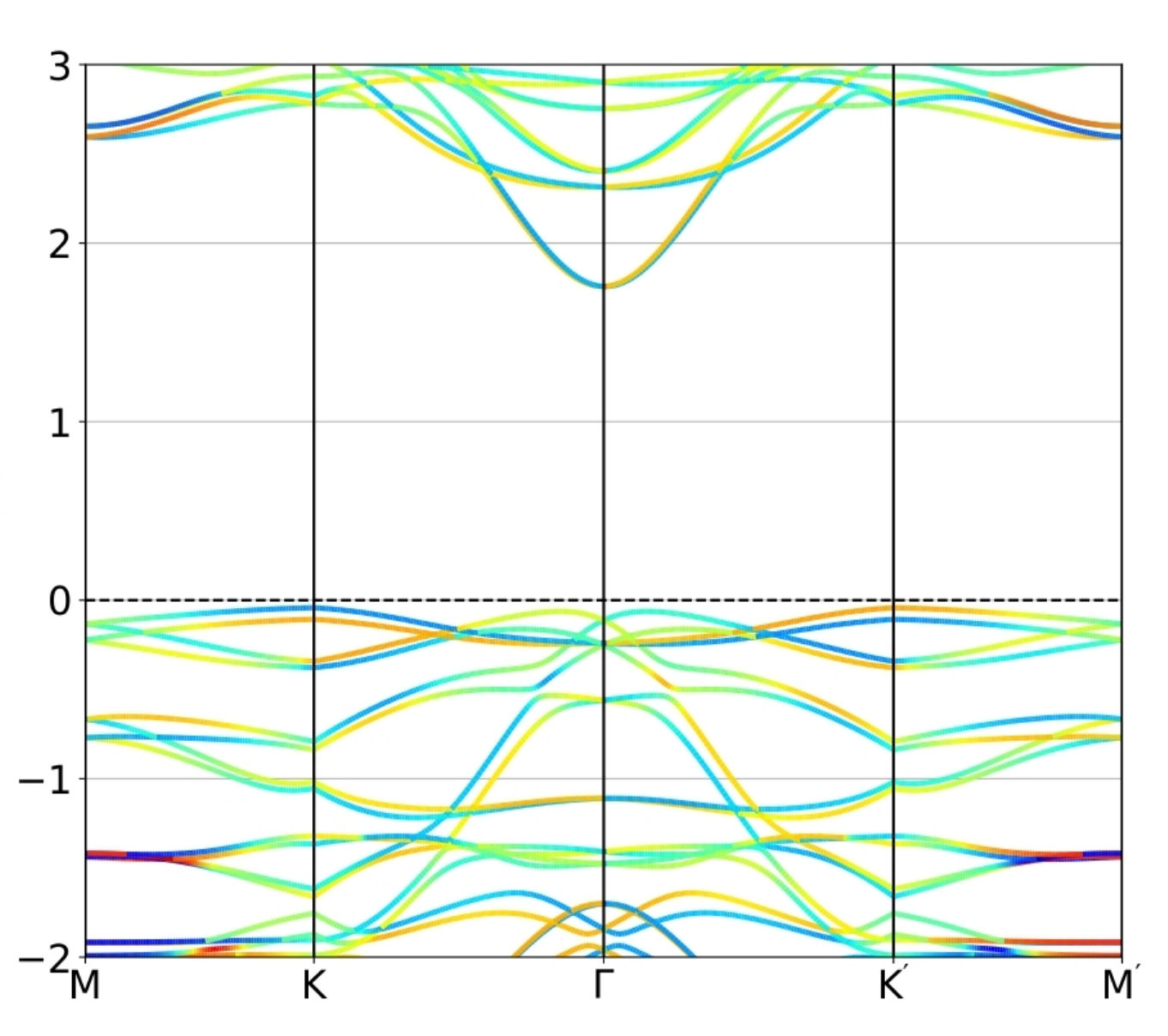}
\includegraphics[scale=0.22]{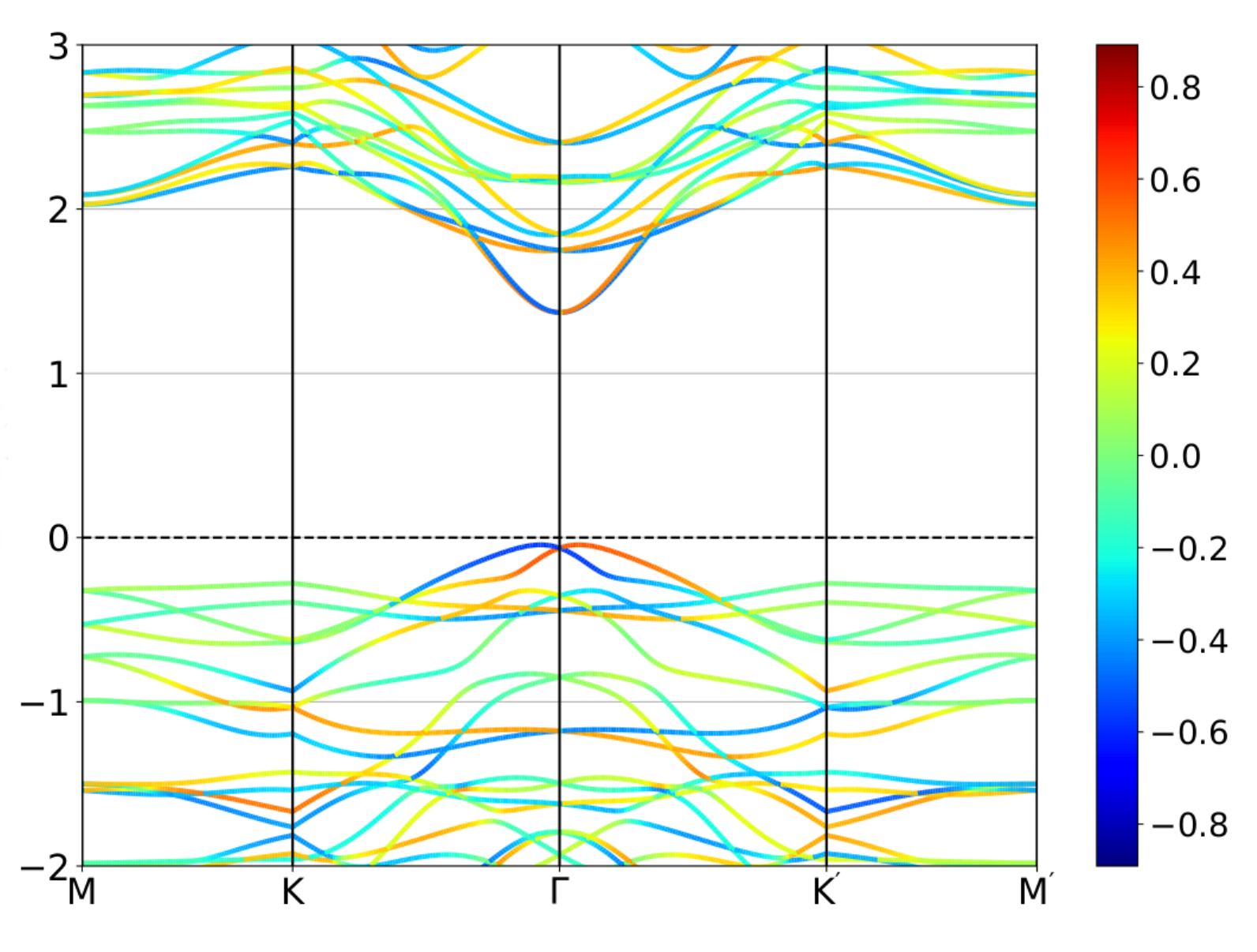}  \\
 \hskip -0.4in  (a) \hskip 1.8in (b) \hskip 1.8in (c)\\
 \vspace{-0.15in}
\caption{(Color online) Band structure plots, with inclusion of SOC, depicting spin-splitting in Sn-doped CdX 
(X = S, Se and Te) monolayers  are shown in (a), (b) and (c) respectively.  The respective Fermi 
energy is set to zero in all plots. (a)~depicts the expectation value of $S_x$ spin component 
(plot for $S_y$ is same with the roles of up and down spin flipped) with the inset showing the enlarged view indicating spin-splitting and
spin-flip at K and K$\textquotesingle$ valleys. 
(b)~depicts the expectation value of $S_y$ spin component (plot for $S_x$ is same with the roles of up and down spins flipped) where the spin-splitting
and spin-flip at K and K$\textquotesingle$ valleys is clearly visible.
(c)~depicts the expectation value of $S_z$ spin component where the
Rashba splitting is clearly visible at $\Gamma$ valley. 
Colors quantify the expectation values of the respective spin components for the bands. 
The extreme values with red and blue colors indicate up and down directions respectively.}
\label{spin-split}

 \vspace{-0.15in}
\end{figure*}

Figure~\ref{spin-split} depicts the band structures of Sn-doped CdX monolayers with inclusion of 
SOC in the calculations. The expectation values of $S_x$ and $S_y$ spin components are same (the roles of up and down spins are flipped) for each monolayer
but the expectation value of $S_z$ spin component is different. Therefore, we have shown the $S_x$, $S_y$ and
$S_z$ components for Sn-doped CdS, CdSe and CdTe respectively.
Inclusion of spin-orbit coupling lifts the degeneracy of both the valence and the conduction bands near the 
valleys and the bands are spin-split, the splitting increases from CdS to CdTe. 
Sn-doped CdS and CdSe 
monolayers have local minima along with spin-splitting at K and K$\textquotesingle$
points in the CB (not visible in Fig.~\ref{spin-split} (a)) but these minima lie higher in energy than 
the CBM at $\Gamma$ point. Further, the splitting is also very small, 
we will therefore not discuss the spin-splitting of the conduction band at the valleys.

The spin-splitting in VB is more than that in CB for all the systems. This is because the 
orbitals participating in the spin-splitting in the valence band region are different from the orbitals in 
the conduction band region. The K and K$\textquotesingle$ valleys show  spin-splitting  
for all the monolayers and presence of time reversal symmetry  flips the spins.
Figures~\ref{band-prestine-soc} and \ref{spin-split} nicely 
depict this split and flip. $\Gamma$ and M points are the time-reversal 
symmetry points and the degeneracy of energy bands at these points is protected by the time-reversal symmetry.
The maxima in VB do occur at K and K$\textquotesingle$ for all the Sn-doped monolayers and 
they are higher than the value of  energy of highest VB at $\Gamma$ point for Sn-doped CdS 
and CdSe monolayers as seen in Fig.~\ref{spin-split}(a) and (b).

The spin-splitting at K and K$\textquotesingle$ valleys is again Zeeman-type for all the  Sn-doped
monolayers like pristine monolayers. The bands are formed by the out-of-plane
Sn $p_z$ states and in-plane $p_y$ and out-of-plane $p_z$ states of X2 and X3 chalcogen atoms, 
as revealed from Table S4 in SI, lifting the spin degeneracy. 
The magnitude of spin-splitting at K and K$\textquotesingle$ points in 
VB for all the systems are tabulated in Table~\ref{ss}.
The spin-splitting in the VB  at K and K$\textquotesingle$ for CdTe is maximum
in the chalcogen series which is quite large and comparable with the reported 
value for MN$_2$X$_2$ (M = Mo, W; X = F, H)~\cite{dou} and MoS$_2$~\cite{korm}.

\begin{table}[b]
\vskip -0,2in 
\caption{Magnitudes of spin-splitting at K and K$\textquotesingle$ points due to
SOC for pristine and Sn-doped CdX monolayers are listed in meV. Values of Rashba parameter $\alpha_R$ at $\Gamma$ point are also listed for Sn-doped CdSe and CdTe
monolayers in eV\AA.}
\begin{center}
\begin{tabular}{ | c | c | c |  c  | }
\hline
  &~~~CdS~~~ &~~CdSe~~~ &~~CdTe~~~\\
\hline
 Pristine~~&11.6 & 16.9 & 48.8\\
 \hline
  Sn-doped~~&     10.3 &64.9&116.5  \\
         \hline   
   $\alpha_R$  & -- & 2.04 &1.8 \\ 
   \hline         
\end{tabular}
\end{center}
\label{ss}
\end{table}
\begin{figure*}
\includegraphics[scale=0.16]{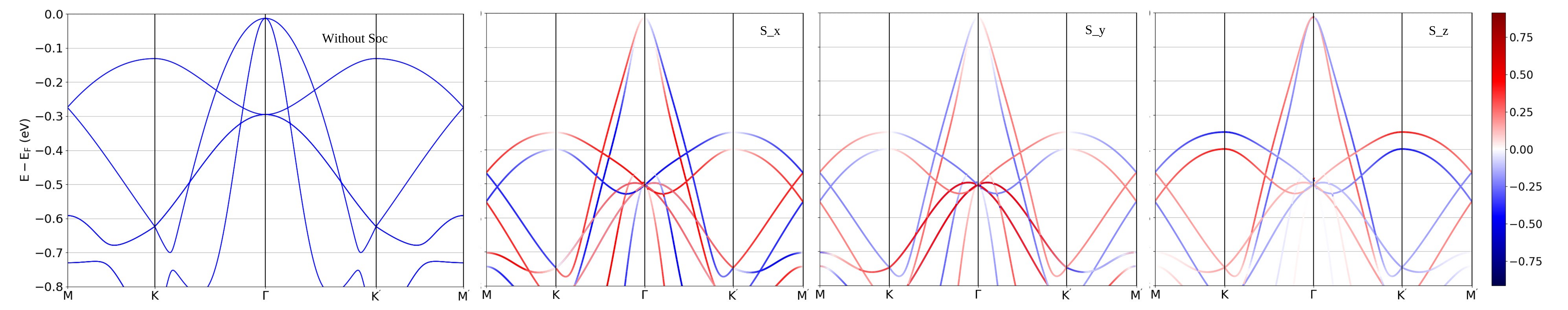}\\
 \hskip 0.2in  (a) \hskip 1.3in (b) \hskip 1.3in (c) \hskip 1.3in (d)\\
 \vspace{-0.15in}
\caption{(Color online) Band structure plots in the vicinity of Fermi level for pristine CdTe monolayer.
The respective Fermi energy is set to zero in all plots. (a) depicts the band structure without inclusion of SOC.
(b), (c) and (d) respectively depict the contributions of S$_x$, S$_y$ and S$_z$ spin components to the bands after inclusion of SOC
in the calculations, with red and blue colors indicating up and down directions respectively.}
\label{CdTe-spin-split}
\vspace{-0.15in}
\end{figure*}

\begin{figure*}
\includegraphics[scale=0.16]{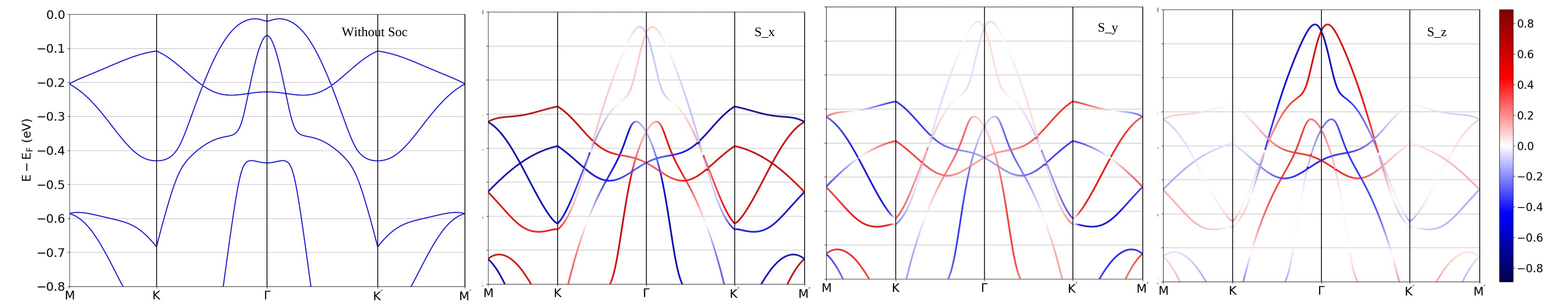}\\
 \hskip 0.2in  (a) \hskip 1.3in (b) \hskip 1.3in (c) \hskip 1.3in (d)\\ 
 \vspace{-0.15in}
\caption{(Color online) Band structure plots in the vicinity of Fermi level for Sn-doped CdTe monolayer. 
The respective Fermi energy is set to zero in all plots. (a) depicts the band structure without inclusion of SOC.
(b), (c) and (d) respectively depict the contributions of $S_x$, $S_y$ and $S_z$ spin components to the bands after inclusion of SOC
in the calculations, with red and blue colors indicating up and down directions respectively.}
\label{CdSnTe-spin-split}
 \vspace{-0.15in}
\end{figure*}

For Sn-doped CdTe monolayer, Rashba spin-splitting is seen at $\Gamma$ point as shown in 
Fig.~\ref{spin-split}(c) and as a result VBM occurs in proximity of $\Gamma$ point. 
Rashba spin-splitting is also seen at $\Gamma$ point for Sn-doped CdSe as shown in 
 Fig.~S6(b) in SI. Rashba spin-splitting
at $\Gamma$ point results  from the splitting of $S_z$ component of the spin,  having the contribution of $p_x$ and $p_y$ states
of atoms X1 and X4, lying in the $xy$-plane, that lifts the  in-plane spin degeneracy.  
It is known that Rashba splitting normally arises at wave vector $\vec k$ with
time-reversal symmetry and therefore is likely to occur at $\Gamma$ point.
The Rashba Hamiltonian
for 2D free electron gas having inversion asymmetry along $\vec z$, is given by 
$H_R=\alpha_R(k \times \vec z) \cdot \sigma$. Here $\sigma$ is the Pauli matrix vector 
and $\alpha_R$ is the Rashba parameter that depends on the atomic number $Z$.
Rough estimates of $\alpha_R$ from DFT ($\alpha_R \approx 2 E_R/k_R$, where $E_R$ is the
change in energy or Rashba energy and $k_R$ is the momentum change or offset) are listed in Table~\ref{ss}
for Sn-doped CdSe and CdTe. Thus, the Rashba spin-splitting lifts the degeneracy
in energy as well as momentum and is dependent on the atomic species.  The values of $\alpha_R$, as mentioned in
Table~\ref{ss}, are comparable with the value of 1.03~eV\AA{} for MoTe$_2$~\cite{gupta}.

To understand the role of SOC and Sn-doping in phenomena important for valley physics, we have plotted the valence bands of 
pristine CdTe and Sn-doped CdTe upto 0.8~eV below the Fermi level in Figs.~\ref{CdTe-spin-split} and \ref{CdSnTe-spin-split} 
respectively. Panel (a) in both the figures depicts the band structures of pristine CdTe and Sn-doped CdTe without the 
inclusion of SOC (spin-up and spin-down bands are identical) while panels (b), (c) and (d) respectively show the 
contributions of $S_x$, $S_y$ and $S_z$ spin components to the bands after inclusion of SOC
in the calculations. Site-projected and $\ell m$-decomposed DOS for Te and Sn atoms in pristine and Sn-doped 
CdTe are plotted in Fig.~S10 in SI. It is observed that SOC increases the hybridization of $p_x$ and $p_y$ 
states of Te with its $p_z$ states and overall the states are broadened in energy extent. Sn-doping enhances 
these features further due to hybridization of Sn and Te $p_z$ states. As a result of hybridization of Te and 
Sn $p_z$ states, the energy degeneracies present in the pristine CdTe band structure at $\Gamma$, K and
K$\textquotesingle$ points are lifted on Sn-doping (compare Figs.~\ref{CdTe-spin-split}(a) and 
\ref{CdSnTe-spin-split}(a)) and the overall nature of band dispersion changes. Inclusion of SOC causes spin-splitting 
of bands as expected. However, the degeneracy of VBM at $\Gamma$ point is retained in pristine CdTe after inclusion of 
SOC, the nature of band dispersion is almost same for up and down spins with spin flipping and more importantly, the 
contributions of $S_x$, $S_y$ and $S_z$ components of spins are almost equally visible in the region of interest. 
Sn-doping significantly changes the band dispersions, more so on inclusion of SOC. The nature of dispersion of up
and down spin bands is different and  $S_z$ component of spin dominates near $\Gamma$ point for VBM and the band below
it whereas the other two spin components contribute almost in the complete BZ below -0.3~eV. This analysis emphasizes 
the role of Sn-doping and inclusion of SOC and further can preempt experiments for valley physics.

The Sn $p$ and X $p$ orbitals hybridize strongly in all Sn-doped CdX monolayers.
After substitution of Cd atoms by Sn atoms, small charge asymmetry arises as
shown in difference charge density plots (Fig.~S9 in SI) and Table S2 of SI.
These observations, on inclusion of
SOC and inversion asymmetry, show enhanced Zeeman-type spin-splitting in
comparison to the respective pristine monolayers at K and K$\textquotesingle$ points and Rashba splitting at $\Gamma$ point in Sn-doped CdSe and CdTe monolayers.
Therefore, all the pristine and Sn-doped systems are
very promising candidates for valleytronics, both to understand the valley physics and for applications.

\begin{figure*}
\includegraphics[scale=0.35]{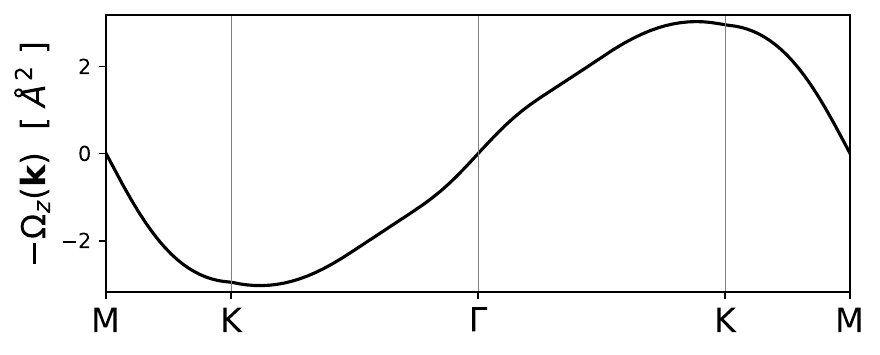}
\includegraphics[scale=0.35]{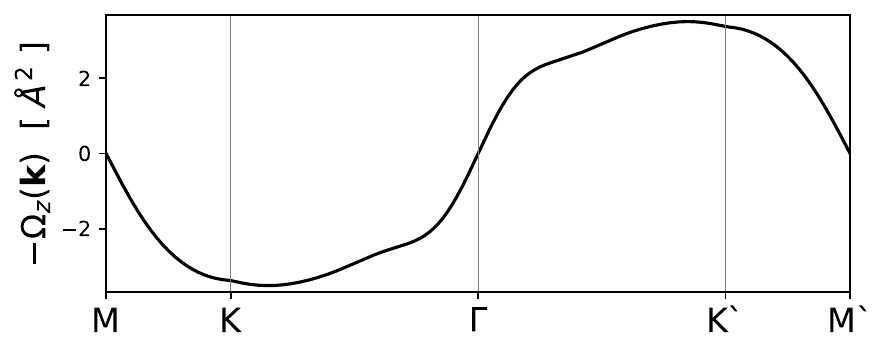}
\includegraphics[scale=0.35]{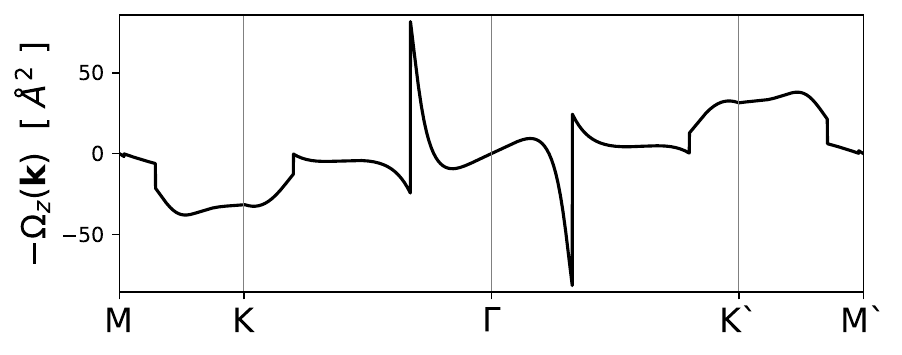}  \\
 \hskip 0.3in (a) \hskip 1.85in (b) \hskip 1.9in (c) \\
\vskip -0.1in
\caption{(Color online) Berry curvature as a curve along the 
high-symmetry points  for Sn-doped  CdX  monolayers is shown 
in (a), (b) and (c) respectively for X = S, Se and Te.}
\label{berry}
\end{figure*}

To validate the valley contrasting character, we have calculated
the Berry curvature of the Sn-doped monolayers. We employed the Kubo 
formulation for the calculation of out-of-plane Berry curvature at K and 
K$\textquotesingle$ valleys as  given by the following equation:
\cite{kubo,yao}.
\begin{equation}
\Omega_z(\vec{k})=-\sum_n\sum_{n\neq n'}f_n\frac{2Im\textless\psi_{n\vec{k}}|\hat{v}_x|
\psi_{n'\vec{k}}\textgreater\textless\psi_{n'\vec{k}}|\hat{v}_y|\psi_{n\vec{k}}\textgreater}{(E_n-E_{n'})^2}
\end{equation}
$\Omega_z(\vec{k})$ is the Berry curvature for all the occupied band at wave vector $\vec{k}$, $f_n$ is the equilibrium Fermi-Dirac
distribution function, $\hat{v}_{x(y)}$ is the velocity operator along $x(y)$ direction, and $\psi_{n\vec{k}}$ is the
Bloch wave function with eigenvalue $E_n$. The summation for each $\vec{k}$ is over all the occupied bands.
Berry curvature plots in Fig.~\ref{berry} show  equal values with 
opposite signs at K and K$\textquotesingle$ points. Thus we can achieve valley 
Hall effect by applying in-plane electric field. The carriers near K and
K$\textquotesingle$ valleys will experience effective magnetic field due to 
the Berry curvatures leading 
to anomalous transverse velocity resulting in valley Hall effect. For all the Sn-doped monolayers, we get inequivalent valleys at K and K$\textquotesingle$ as the Berry curvature has equal magnitude but opposite signs. The Berry
curvature varies inversely with $k$ and with the band gap~\cite{raihan}. Therefore, the Berry curvature values are not very large for Sn-doped CdS and CdSe monolayers but for Sn-doped CdTe monolayer,
we get spikes, with very large values, in the Berry curvature plot between K
(K$\textquotesingle$) and
$\Gamma$ points, in proximity of $\Gamma$ point.

These spikes are a result of avoided crossing of the hybridized bands that are spin-split due to spin-orbit interaction and orbital hybridization. Sn-doped CdTe, thus, can be a potential candidate for valleytronics, more so with thicker layer and reduced band gap. The Berry curvature values are small for pristine monolayers as compared to Sn-doped monolayers.  In case of Sn-doped CdS monolayer, the Berry curvature plot is almost smooth as the SOC effect is very weak and the spin-splitting of bands is small in value. The plot for Sn-doped CdSe does show small kinks near the $\Gamma$ point due to sizable spin-splitting and spin-flip.
Our values of Berry curvature of Sn-doped CdTe monolayer are comparable with those reported by Jena~\textit{et al.}~\cite{jena} for MoS$_2$ and
by Raihan~\textit{et al.}~\cite{raihan} for NbN and TaN.

\section{Conclusions}
In this study, we have designed a new class of monolayers by doping easily 
available non-toxic Sn atoms in hexagonal CdX (X = S, Se, Te) monolayers. Our DFT
calculations depict two valley points, K and K$\textquotesingle$,
with valley spin-splitting in both the VB and
CB bands with the application of SOC. The magnitude of valley spin-splitting
in VB is large in comparison to CB. Sn atoms do enhance
the magnitude of valley spin-splitting.
Valleys show Zeeman-type spin-splitting in all the three Sn-doped monolayers while in case of Sn-doped CdSe
and CdTe monolayers, Rashba spin-splitting is observed at $\Gamma$ point 
which is extremely important for valleytronic and spintronic applications.
Spin-valley locking with Zeeman-type splitting and Rashba spin-splitting, due to SOC, are the most flourishing
fields of study in valleytronics and spintronics and these properties can be easily
harnessed in the semiconducting CdX monolayers by doping environmentally safe and earth-abundant Sn atoms.

Structurally, pristine and Sn-doped CdS monolayers are the most stable but at the same time, the formation energies are also the largest for these monolayers. On the other hand, pristine and Sn-doped CdTe monolayers seem to be energetically realizable from the constituents and due to high Z values experience strong SOC effects resulting in excellent valleytronics properties. Pristine and Sn-doped CdSe also displays good valleytronics properties. Further, thin films and nanosheets of CdSe and  nanocrystalline thin films of Sn-doped CdSe have already been synthesized. We therefore foresee good chances of Sn-doped CdSe and CdTe monolayers/nanosheets to be useful in valleytronic and spintronic devices.

This study suggests
that theoretically predicted 2D Sn-doped CdX hexagonal monolayers reveal an
ideal platform for studying valley physics, Rashba physics,
and their integration in valleytronics with spintronics. Our calculations further demonstrate that
Sn-doped CdTe can be experimentally synthesized using bottom-up approach.

\section{Acknowledgements} SC would like to acknowledge financial support from
Department of Science and Technology, Government of India through their
Women Scientist-A (WOS-A) program. Param Brahma super computer under 
National Supercomputing Mission established at IISER, Pune  has been used for 
the DFT calculations. This support is gratefully acknowledged.
Authors thank Prof. P. Durganandini and Dr. Vikas Kashid for useful discussions.

\section*{References}

\bibliography{ref1}

\end{document}


\vspace{-0.3in}
\title{Design of Sn-doped cadmium chalcogenide based monolayers for valleytronics properties}

\vspace{-0.1in}

\vskip -0.8in

\section{Supplementary Material}

\begin{table*}[h]

{\small{{\bf Table S1.} Lattice parameters ($a = b$) of pristine CdX monolayers along with the
buckling index $d_z$, Cd-X
bond lengths and band gaps $E_g$ (I : indirect and D : direct). 
Values in the brackets are from earlier published works.}}

\centering{
\vspace{0.1in}
\begin{tabular}{ | c | c | c | c | }
\hline
Properties $\downarrow$&  \multicolumn{3}{|c|}{Pristine CdX monolayers} \\
 \hline
  & CdS & CdSe & CdTe\\
\hline
$a(= b)$& 4.11 (4.24$^{a,g}$, 4.23$^b$,& 4.43 (4.40$^b$, 4.32$^{c,d}$,& 4.72 (4.59$^{c,d}$,\\
(in \AA )& 4.16$^{c,d}$, 4.20$^f$, 4.27$^h$)& 4.39$^{e,f}$, 4.44$^{g,h}$)& 4.76$^g$, 4.73$^h$)\\
\hline
$d_z$& 0.06 (0.04$^a$, 0.0$^{c,e,g}$& 0.03 (0.0$^g$, 0.36$^h$, 0.41$^b$,& 0.04 (0.0$^g$, 0.46$^c$,\\
(in \AA )& 0.05$^h$, 0.07$^b$, 0.45$^f$)& 0.31$^c$, 0.32$^e$, 0.68$^f$)& 0.49$^{e,h}$)\\
\hline
Cd-X & 2.52(2.44$^{a,b}$, 2.40$^c$,& 2.58(2.57$^b$, 2.51$^c$,& 2.77(2.69$^c$, 2.77$^h$)\\
(in \AA )& 2.46$^h$, 2.56$^d$, 2.53$^f$)& 2.59$^h$, 2.67$^f$)& \\
\hline
$E_g$ without& 1.65(D) (1.60(D)$^{a,c}$,& 1.26(D), (1.31(D)$^b$,& 1.21(D)(1.24(D)$^c$,\\
 inclusion of& 1.66(D)$^b$, 1.64(D)$^e$,& 1.20(D)$^c$, 1.23(D)$^e$,& 1.20(D)$^e$, 0.59(I)$^g$)\\
SOC (in eV)&  1.58(D)$^f$, 1.55(D)$^g$)& 1.33(D)$^f$, 0.91(I)$^g$&  \\
\hline
$E_g$ with SOC (in eV)& 1.64(D) & 1.13(D) & 0.98(I) \\
\hline 
\end{tabular}}
\end{table*}
\vspace{-0.2in}
\noindent{\small{a:~Ref.~\cite{pgarg}, b:~Ref.~\cite{kishore}, c:~Ref.~\cite{zheng}, d:~Ref.~\cite{opoku}, 
e:~Ref.\cite{zhou}, f:~Ref~\cite{mohanta}, g:~Ref.~\cite{safari}, h:~Ref.~\cite{wang1}.}}

\noindent {\small{The quoted results from previous works, given in brackets, are for similar structures and GGA-PBE functional only.}}

\begin{table*}[h]
\small{{\bf Table S2.} Lattice parameters ($a = b$) of Sn-doped CdX monolayers along with the buckling index $d_z$, bond lengths Cd-X and Sn-X and band gaps $E_g$ (I : indirect and D : direct).}
\vspace{0.1in}
\centering{
\begin{tabular}{ | c | c | c | c | }
\hline
 Properties $\downarrow$ &   \multicolumn{3}{c|}{Sn-doped CdX monolayers}\\
    \hline
  & CdS & CdSe & CdTe\\
\hline
$a(= b)$ (in \AA ) &  3.99 & 4.16 & 4.45 \\
 \hline
$d_z$ (in \AA ) & 1.25 & 1.26  & 1.29 \\
\hline
Cd-X (in \AA ) &  2.47 & 2.78 & 2.78\\
\hline
Sn-X (in \AA ) & 2.58, 2.64 & 2.72, 2.78 & 2.90, 2.98 \\
\hline
$E_g$ without SOC (in eV)  & 2.19 (I)  &  1.85 (I)  & 1.60 (I)\\
\hline
 $E_g$ with SOC (in eV) & 2.15 (I) &  1.80 (I) & 1.41 (I)\\
\hline
\end{tabular}}
\end{table*}

\begin{figure*}
\includegraphics[scale=0.4]{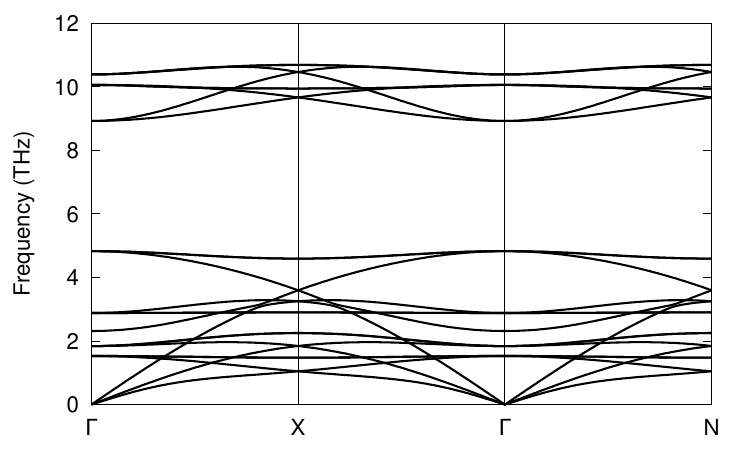}
\includegraphics[scale=0.4]{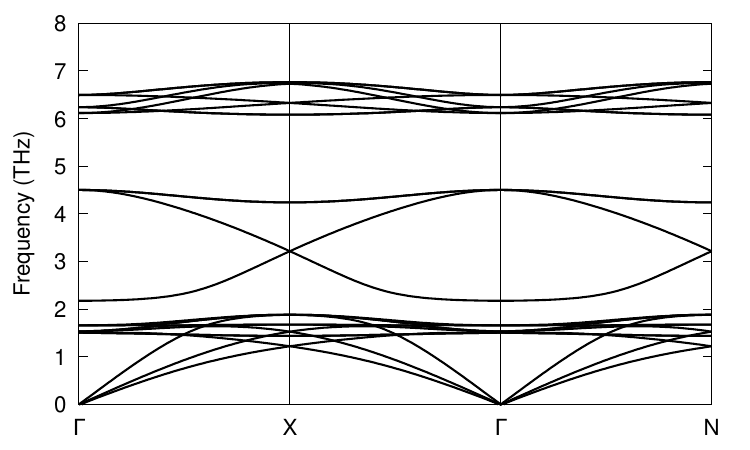}
\includegraphics[scale=0.4]{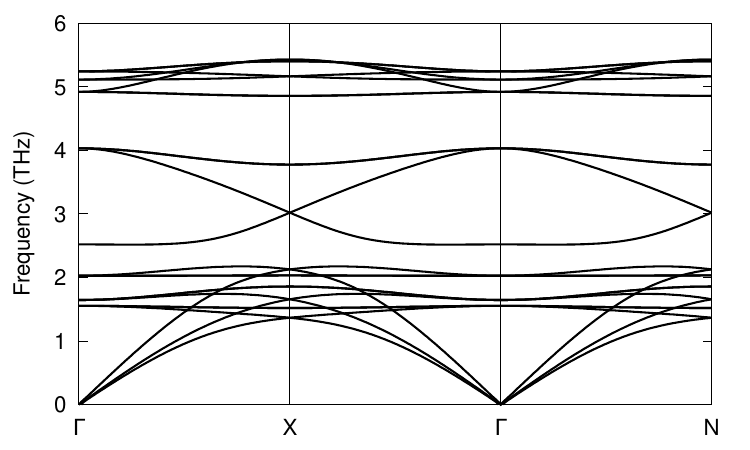}\\
\vspace{-0.1in}
(a) \hspace{1.7in} (b) \hspace{1.7in} (c) \\
\vspace{0.1in}
\includegraphics[scale=0.4]{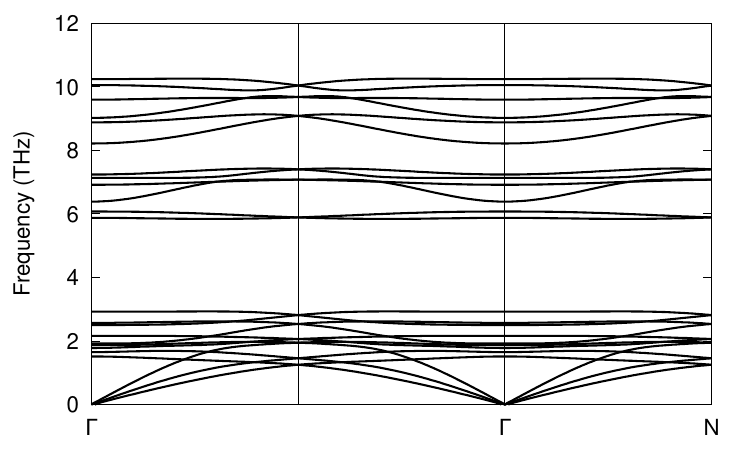}
\includegraphics[scale=0.4]{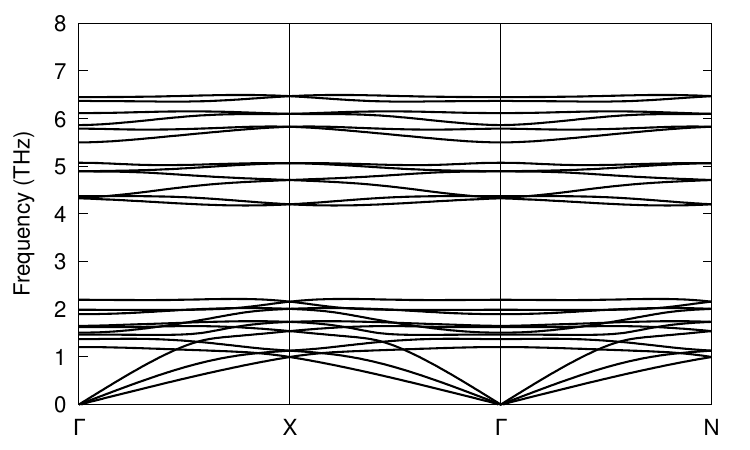}
\includegraphics[scale=0.4]{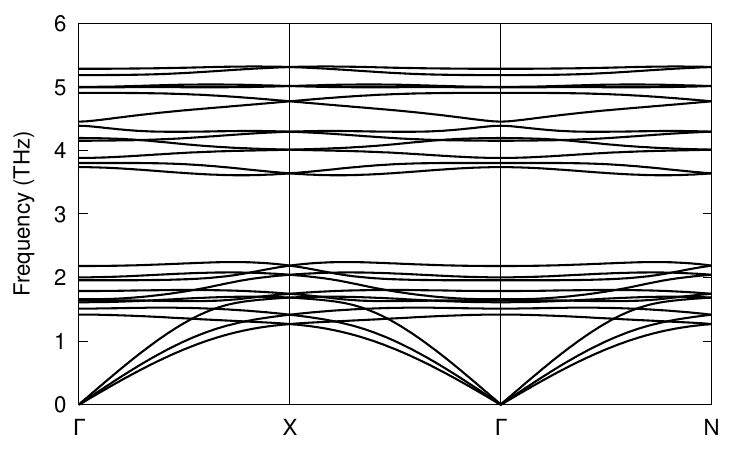}\\
\vspace{-0.1in}
(d) \hspace{1.7in} (e) \hspace{1.7in} (f) \\
{\small{{\bf Figure S1.} Phonon dispersion plots of pristine and Sn-doped CdS, CdSe and CdTe monolayers are shown in (a), (b) and (c) and (d), (e) and (f) respectively. None of these plots shows any negative frequency/imaginary modes establishing  the dynamic stability of all the monolayers studied in this work. The acoustic modes, which are mainly contributed from in-phase vibrations and therefore dominated by heavier atoms (Cd and Sn), mix with the optical modes, resulting from out of phase motion of lighter X (S and Se) atoms about the relatively fixed heavier atoms, away from the $\Gamma$ point.}}
\vspace{-0.1in}
\end{figure*}

\vspace{0.2in}
\begin{figure*}
\vspace{-0.1in}
\includegraphics[scale=0.59]{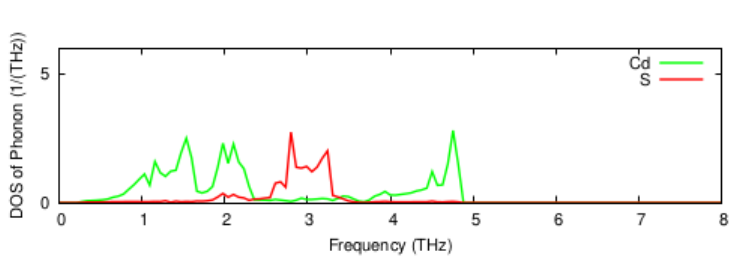}
\includegraphics[scale=0.59]{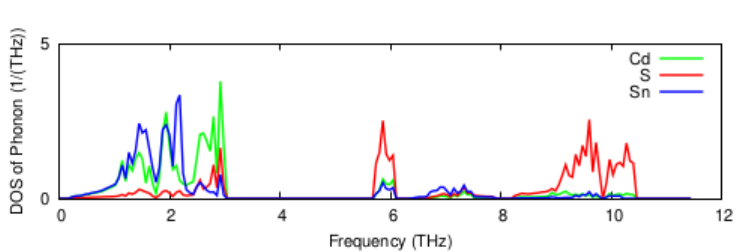}\\
(a) \hspace{2.65in} (b) \\
\includegraphics[scale=0.59]{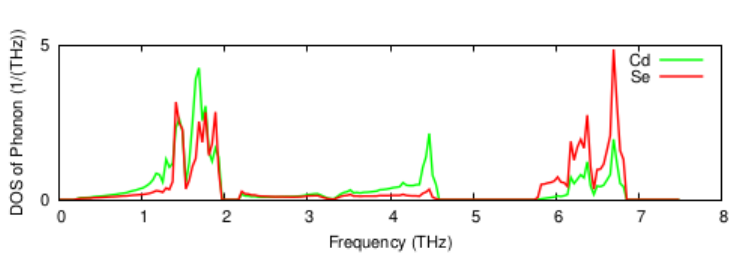}
\includegraphics[scale=0.59]{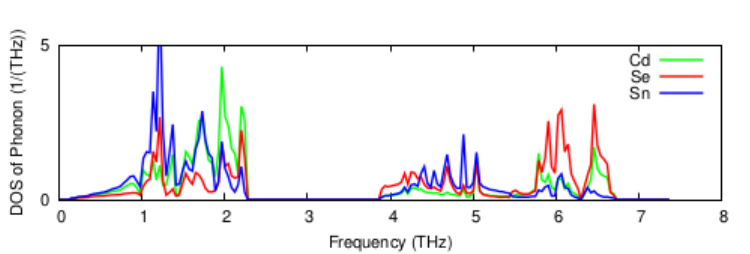}\\
(c) \hspace{2.65in} (d) \\

\includegraphics[scale=0.59]{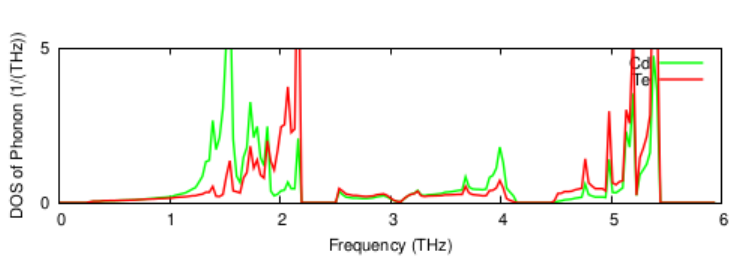}
\includegraphics[scale=0.59]{S2_fig_2e.pdf}\\
(e) \hspace{2.65in} (f) \\
{\small{{\bf Figure S2.} (Color online)  Phonon density of states of pristine and Sn-doped CdS, CdSe and CdTe
monolayers are shown in (a), (c) and (e) and (b), (d) and (f) respectively. 
Red, green and blue curves indicate the contribution of X, Cd and Sn atoms respectively.}}
\end{figure*}

\begin{figure*}
\begin{centering} 
.\hspace{-0.1in}\includegraphics[scale=0.28]{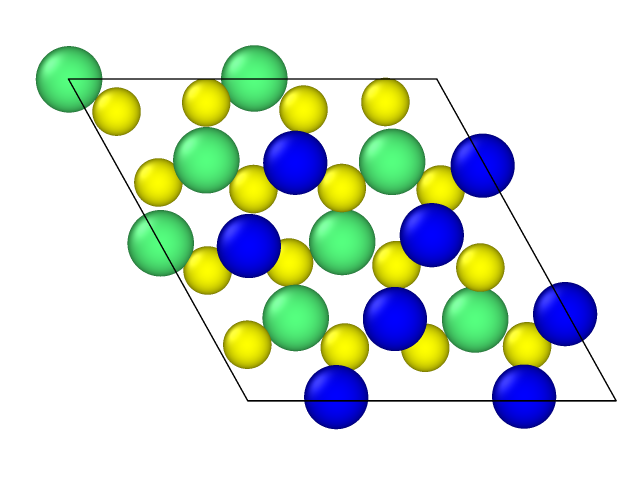}
.\hspace{0.1in}\includegraphics[scale=0.28]{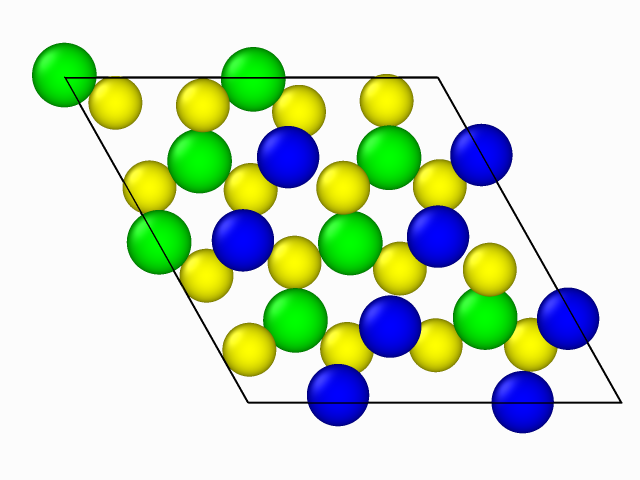}
.\hspace{0.1in}\includegraphics[scale=0.28]{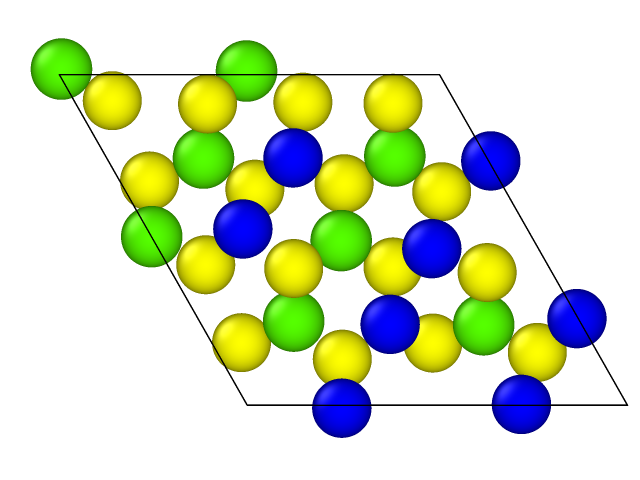}\\
 \includegraphics[scale=0.32]{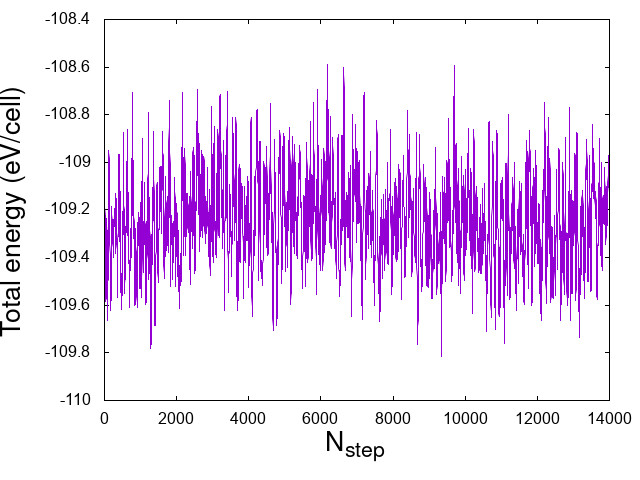}\includegraphics[scale=0.32]
 {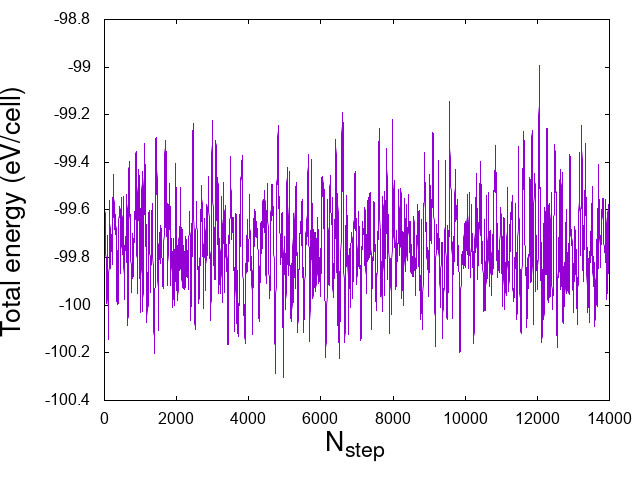}\includegraphics[scale=0.32]{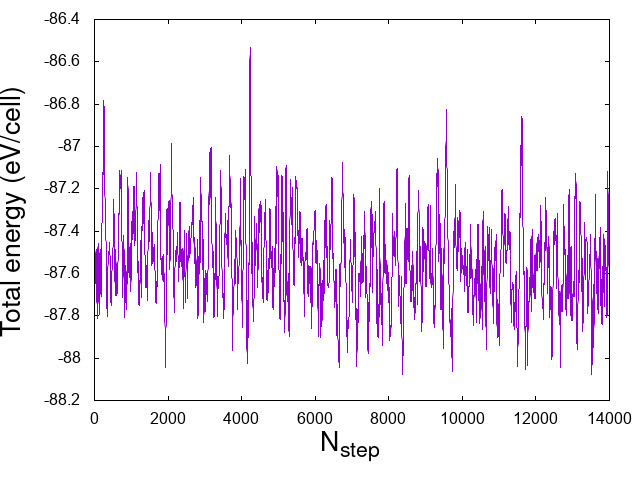}\\
 \end{centering}
  (a) \hspace {1.9in} (b) \hspace {1.8in} (c) 

\begin{centering}
 \vspace{0.2in}
 \includegraphics[scale=0.32]{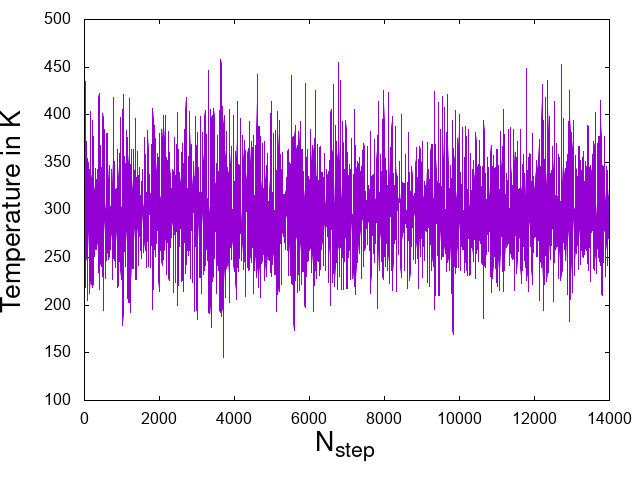}\includegraphics[scale=0.32]
 {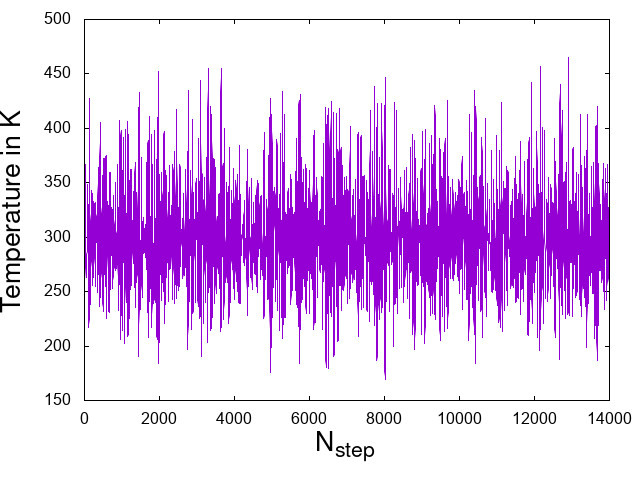}\includegraphics[scale=0.32]{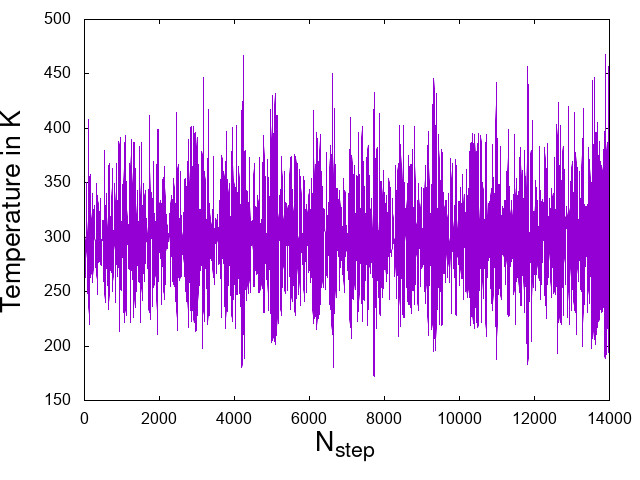}\\
 \end{centering}
  (d) \hspace {1.9in} (e) \hspace {1.8in} (f) \\
  \vspace{0.2in}  
{\small{{\bf Figure S3.} (Color online)  Variation in total energy, using  NVT ensemble, as a function of the time steps, at 300 K in
the MD simulation for Sn-doped CdX (X = S, Se and Te) monolayers is 
shown in (a) (b) and (c) respectively. Corresponding temperature evolution as a function of time steps is shown  in (d), (e) and (f) respectively. The mean temperature is close to 300 K as seen from the plots.
Results of first 1000 time steps are not included in all the plots. 
The average total energies are -109.3~eV, -99.8~eV and -87.6~eV at 300~K from the above plots whereas the total energies from the {\it {ab initio}} calculations are -114.16~eV, -104.02~eV and -92.76~eV at 0~K for Sn-doped CdS, CdSe and CdTe respectively. It may be mentioned that 
total energy for CdS by AIMD has been shown to be -90.0~eV and -88.7~eV at 200~K and 300~K respectively by Garg~\textit{et al.} whereas total energy of CdS from {\it {ab initio}} calculations is -92.18~eV~\cite{pgarg}. Snapshots of the geometries of final step of the MD simulations for
Sn-doped CdS, CdSe and CdTe monolayers are shown at the top.}}
\end{figure*}

\begin{figure*}
\begin{centering}
\includegraphics[scale=0.18]{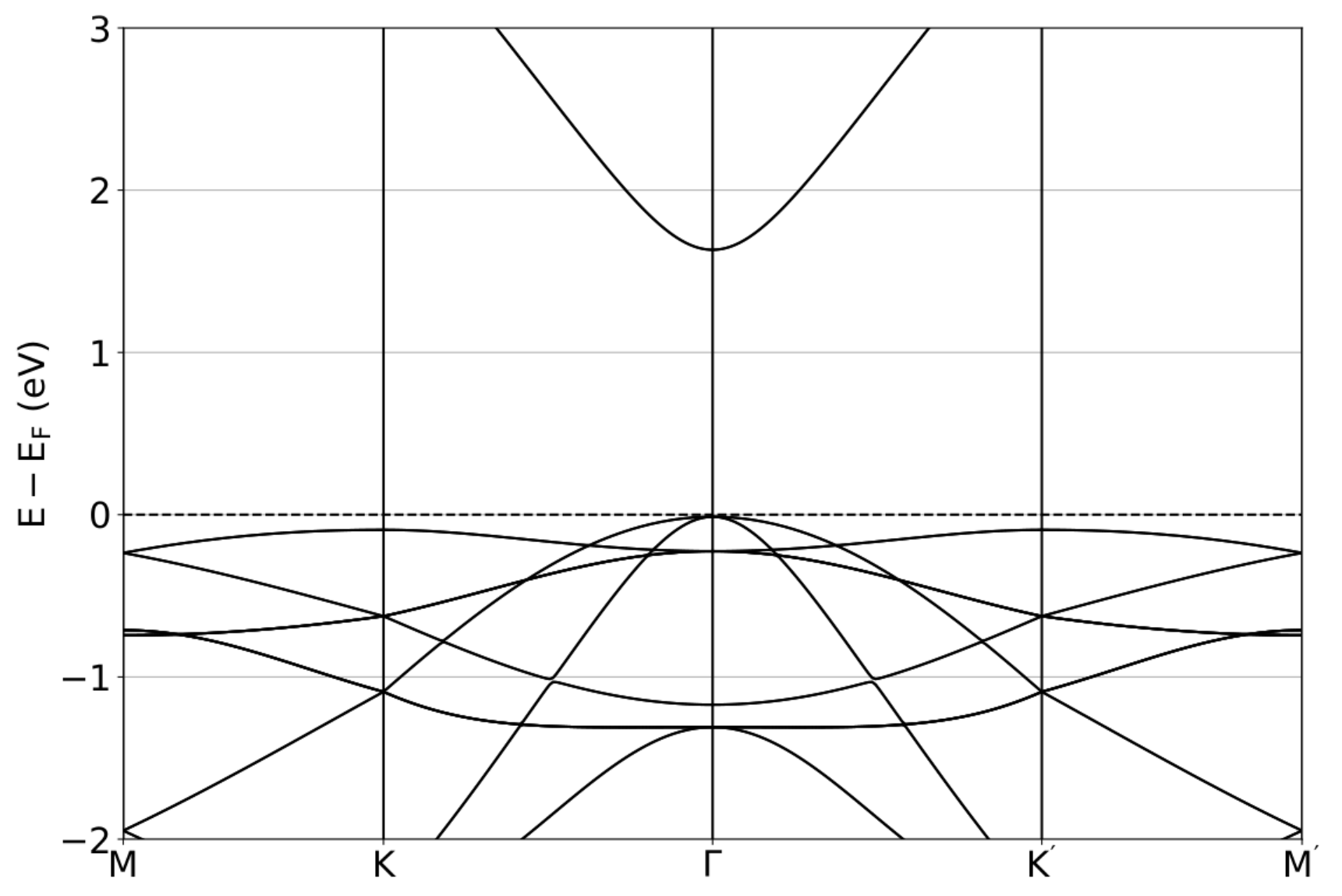}\includegraphics[scale=0.18]
{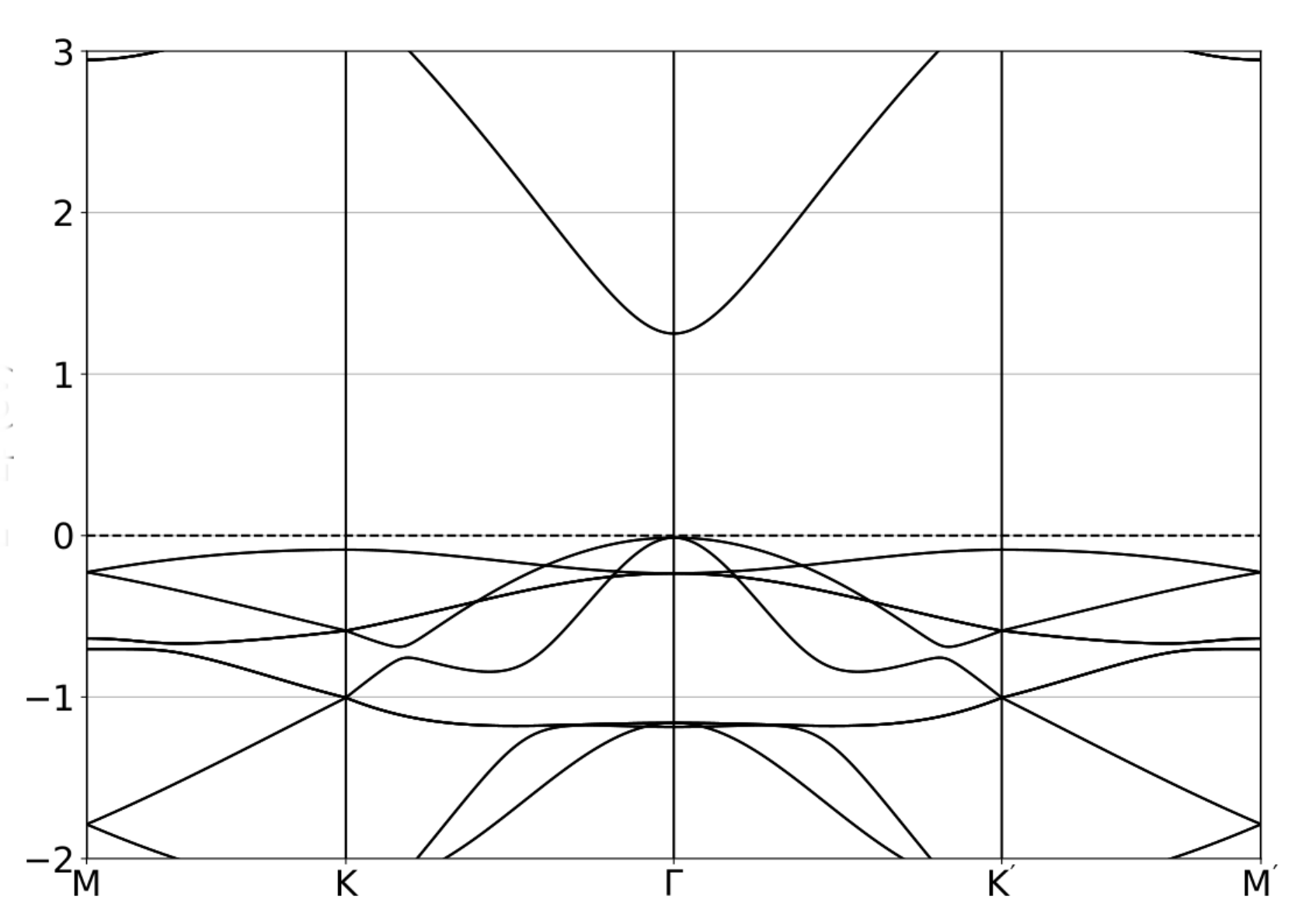}\includegraphics[scale=0.18]{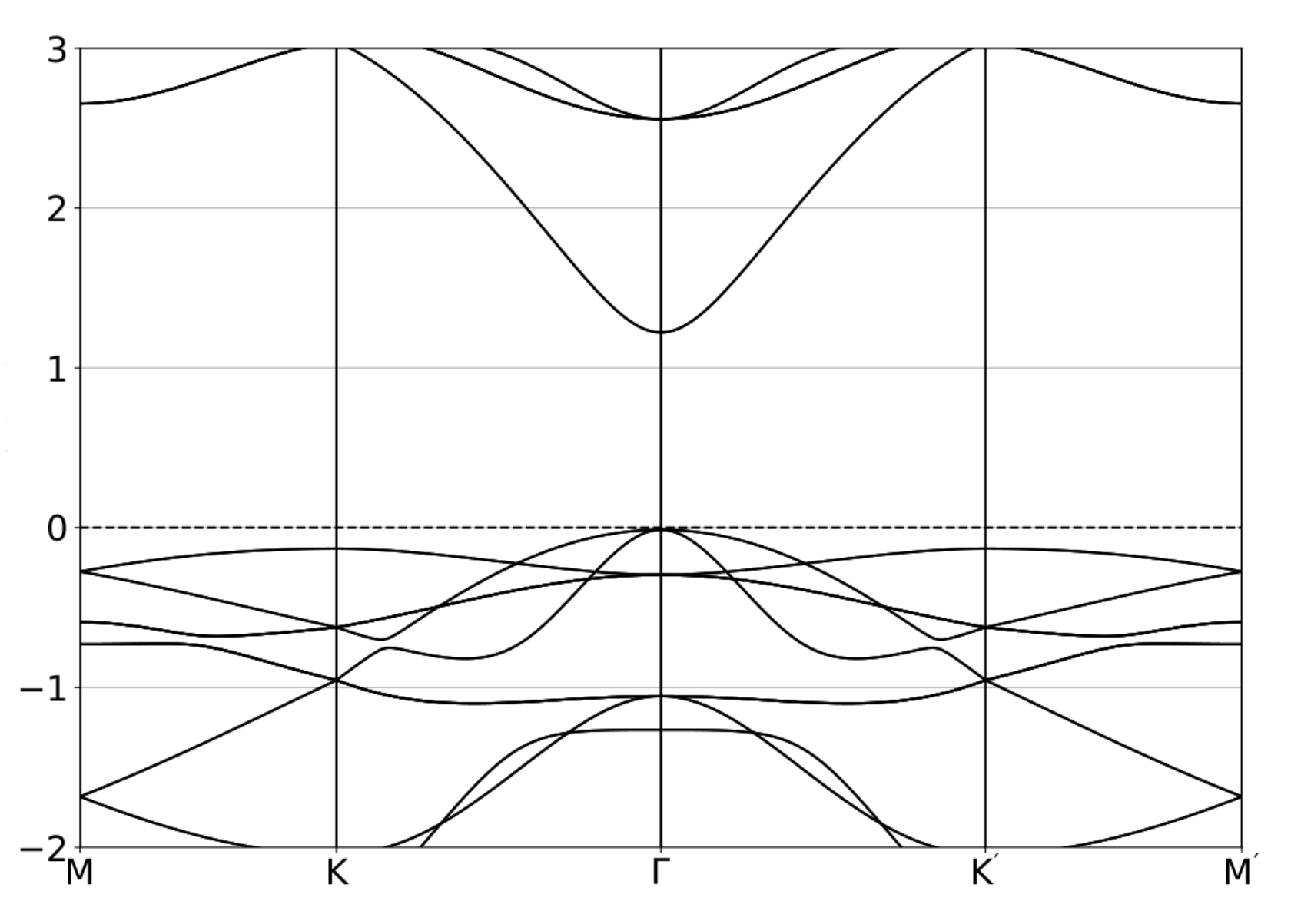}\\
\end{centering}
 (a) \hskip 1.80in (b) \hskip 1.7in (c)
  
  \vskip 0.06in
{\small{{\bf Figure S4.} Band structure plots for pristine CdX (X = S, Se and Te) monolayers, without inclusion of SOC, are
shown in (a), (b) and (c) respectively. 
The respective Fermi energy is set to zero in all the plots. Gradual changes in the bands
arising out of the chalcogen atoms in the valence band region, in particular, 
lifting of degeneracy of one pair of bands at M point, is seen as one moves from S to Te. 
The conduction band minimum is more dispersive and moves downwards from S to Te, reducing the band gap.
Some of the avoided crossings (not all) are visible in (b) and (c).}}
\end{figure*}

\begin{figure*}
 
\includegraphics[scale=0.3]{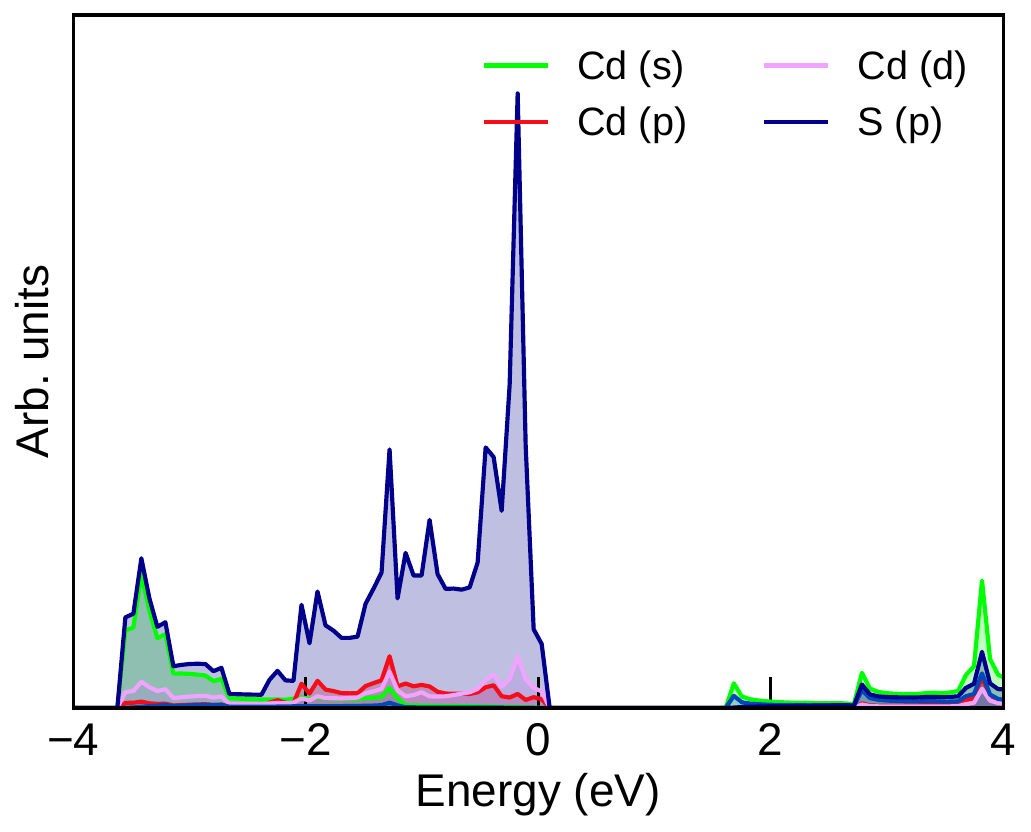}
\includegraphics[scale=0.3]{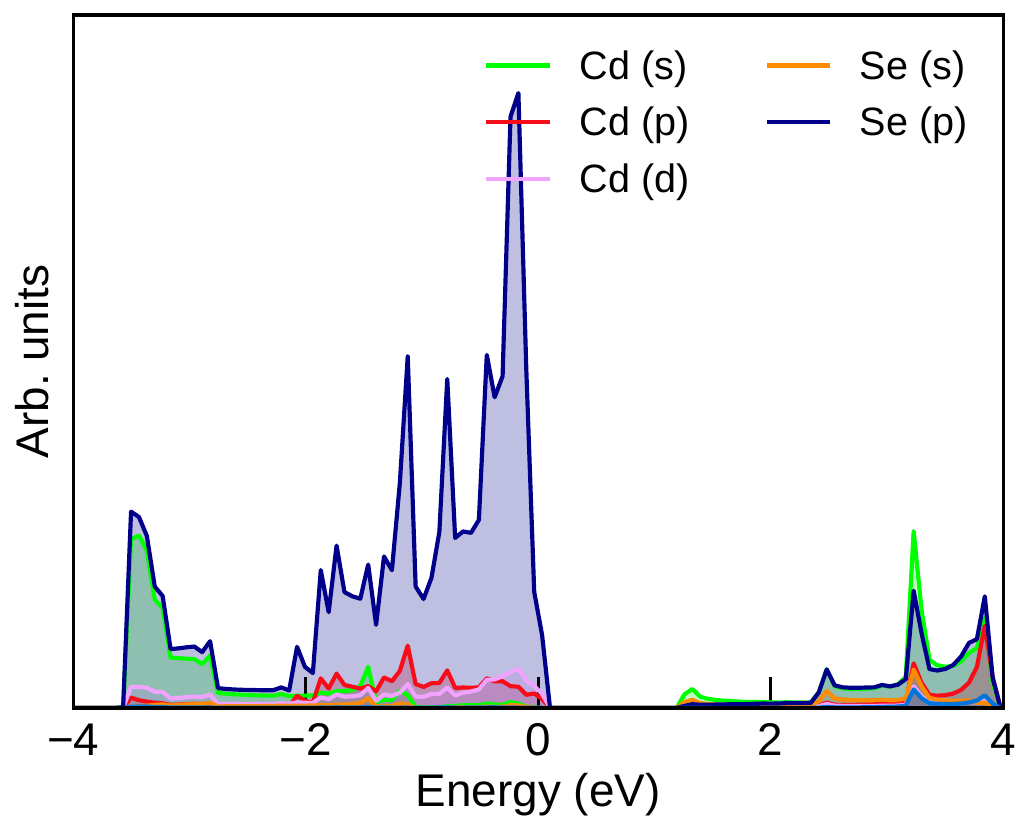}
\includegraphics[scale=0.3]{S_fig5_b.pdf}\\
\vspace{-0.1in}
   (a) \hskip 1.9in (b) \hskip 1.9in (c) \\
\vskip 0.05in
\includegraphics[scale=0.3]{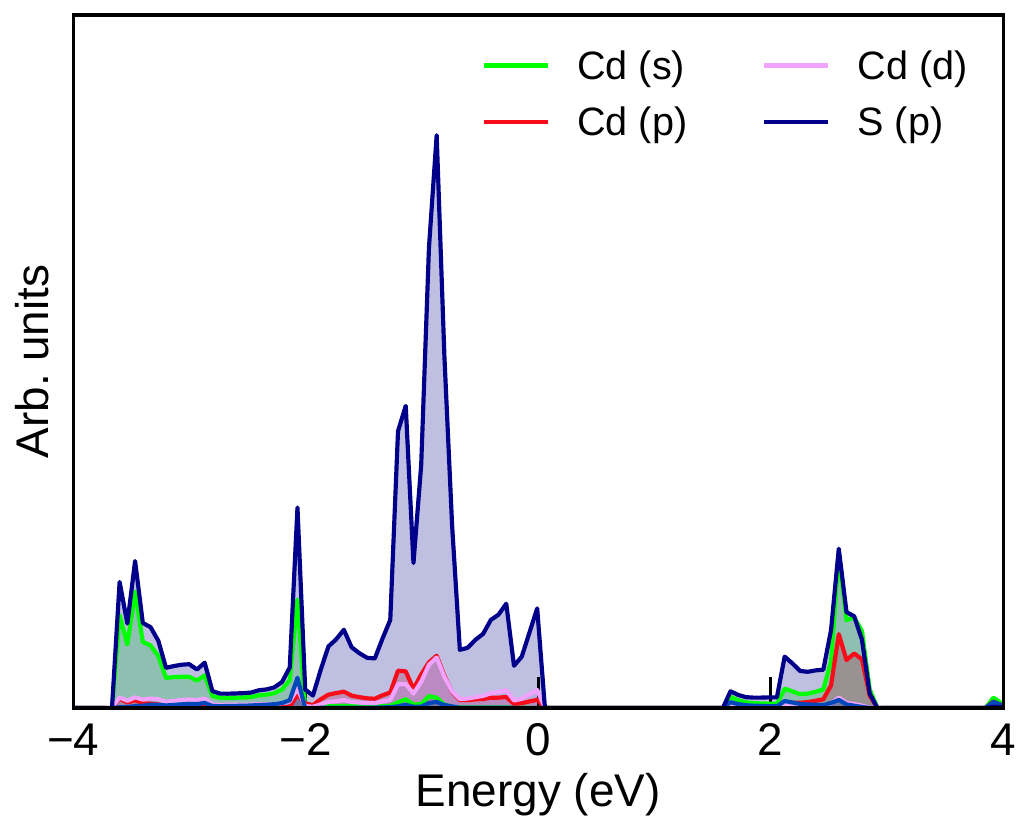}
\includegraphics[scale=0.3]{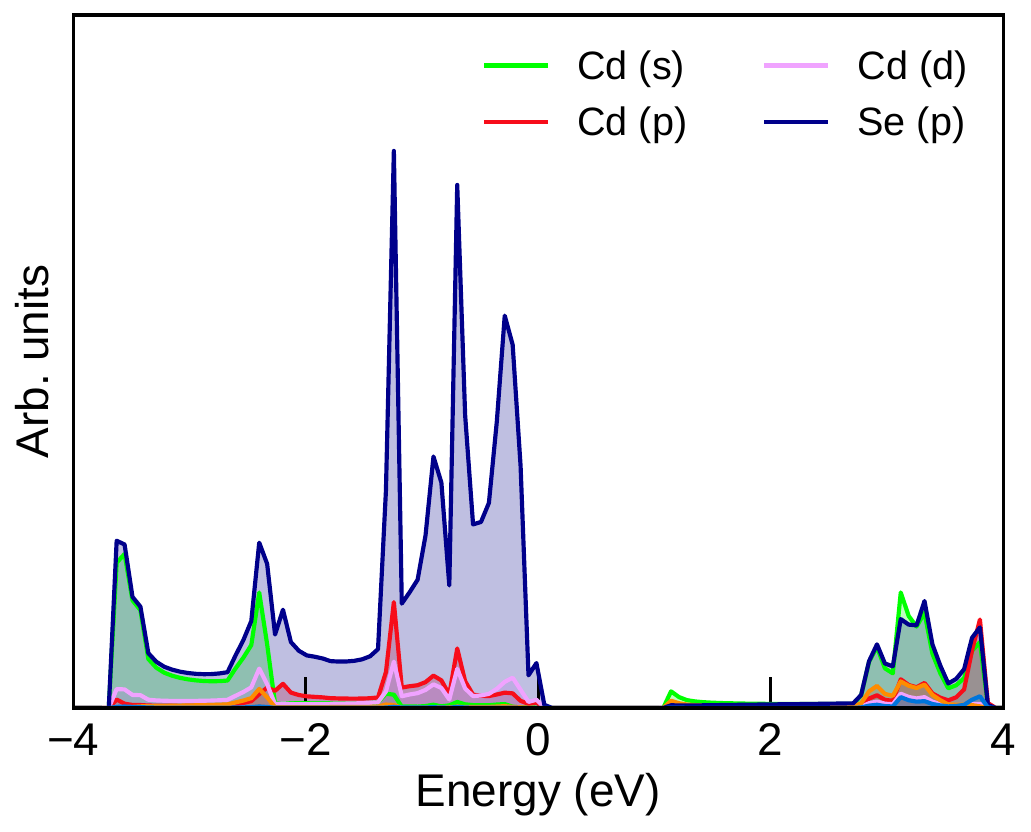}
\includegraphics[scale=0.3]{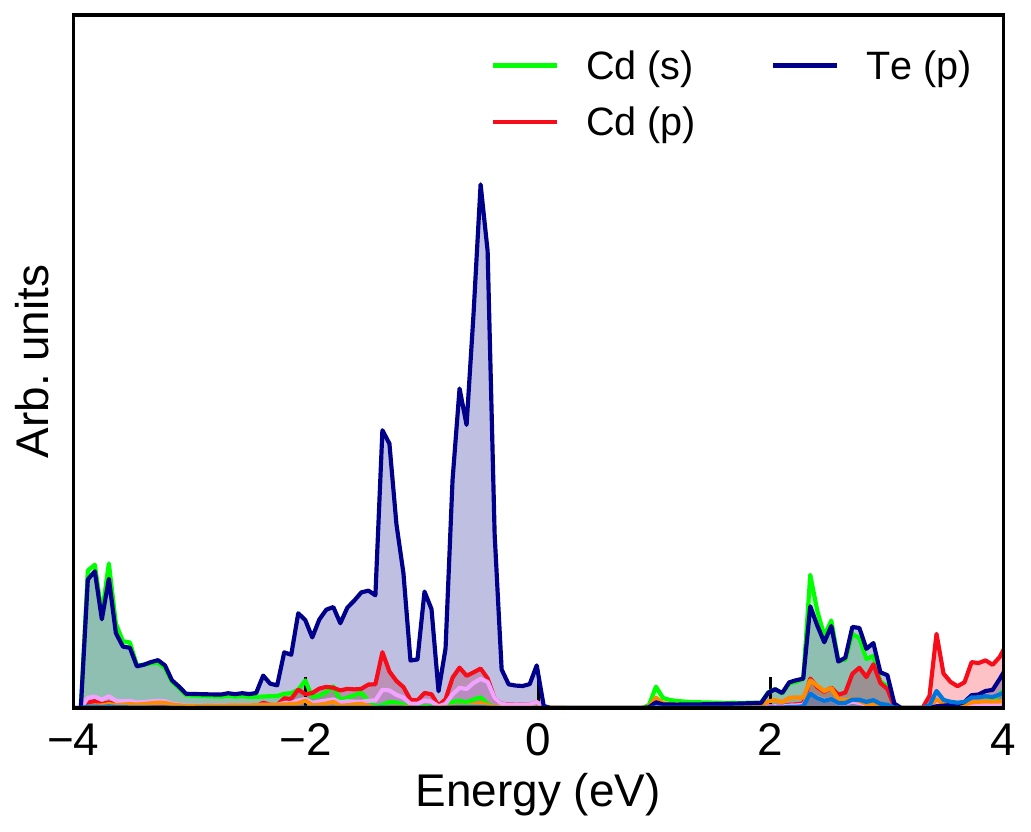}\\
\vspace{-0.1in}
 (d) \hskip 1.9in (e) \hskip 1.9in (f) \\

\vspace{-0.08in}
{\small{{\bf Figure S5.} (Color online) Projected density of states (PDOS) plots of pristine CdX (X = S, Se and Te)
monolayers without inclusion of SOC are shown in (a), (b) and (c) respectively. (d), (e) and (f) depict the same with inclusion of SOC. 
In all the plots states near Fermi energy are dominantly contributed by X $p$ orbitals.}}
\end{figure*}

\begin{figure*}
\begin{centering}
\includegraphics[scale=0.2]{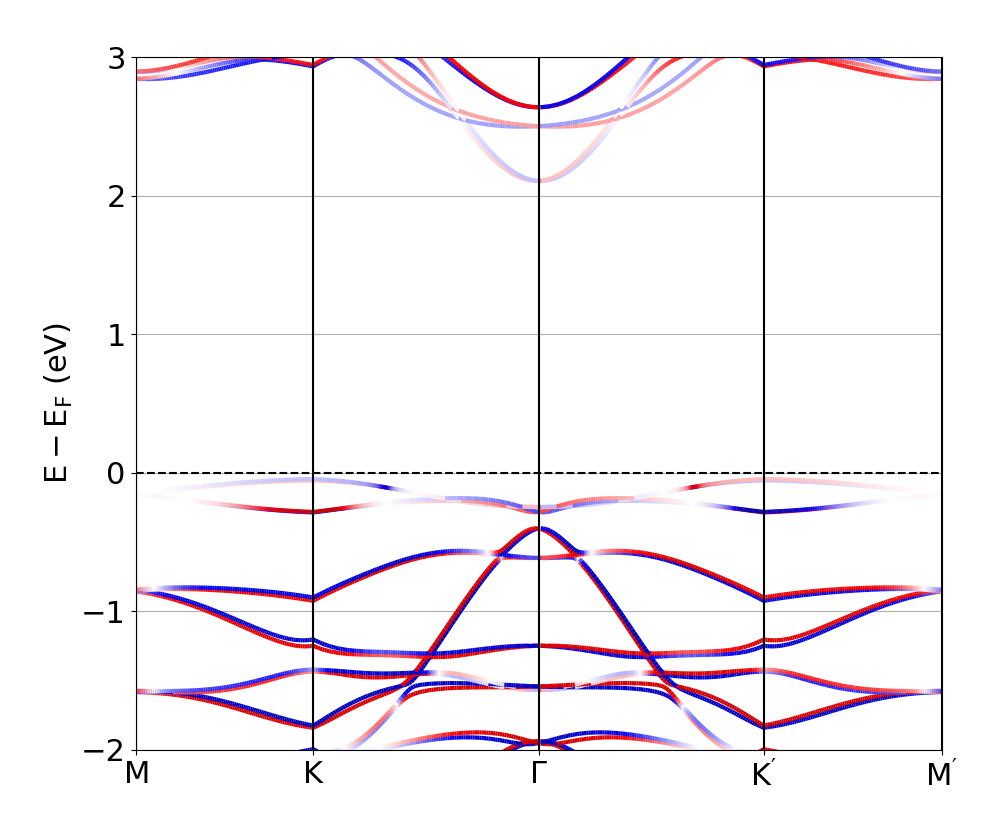}
\includegraphics[scale=0.2]{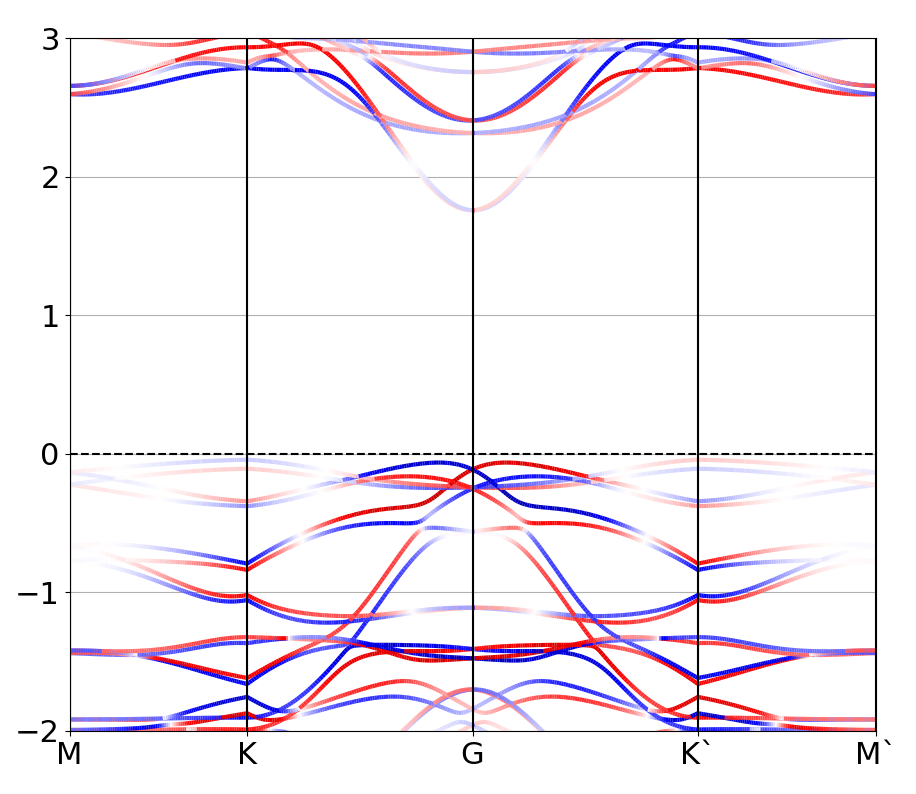}
\includegraphics[scale=0.2]{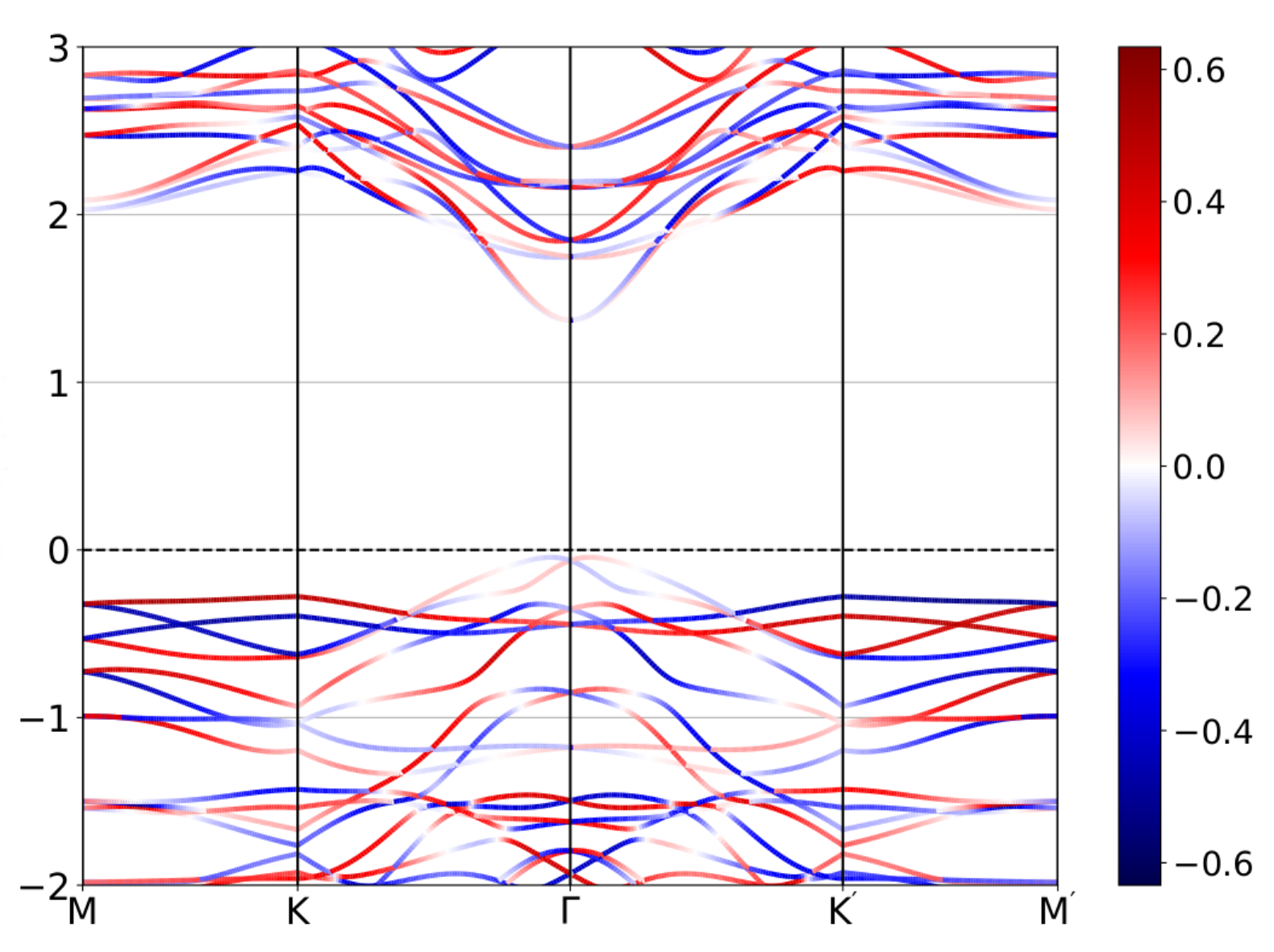}\\
\end{centering}

 (a) \hskip 1.7in (b) \hskip 1.7in (c) 

{\small{{\bf Figure S6.} (Color online) Band structure plots for Sn-doped CdS and CdSe monolayers, with inclusion of SOC,
for the $S_z$ spin component are shown in (a) and (b) respectively. (c) depicts the same for the 
$S_x$ spin component (S$_y$ component is identical with the roles of up and down spins flipped) for CdTe. The respective Fermi energy is set to zero in all the plots.
Red and blue curves represent bands for up and down spins respectively. Some portions
of bands are faint since the particular component of spin does not contribute.}}
\end{figure*}

\begin{figure*}
\includegraphics[scale=0.3]{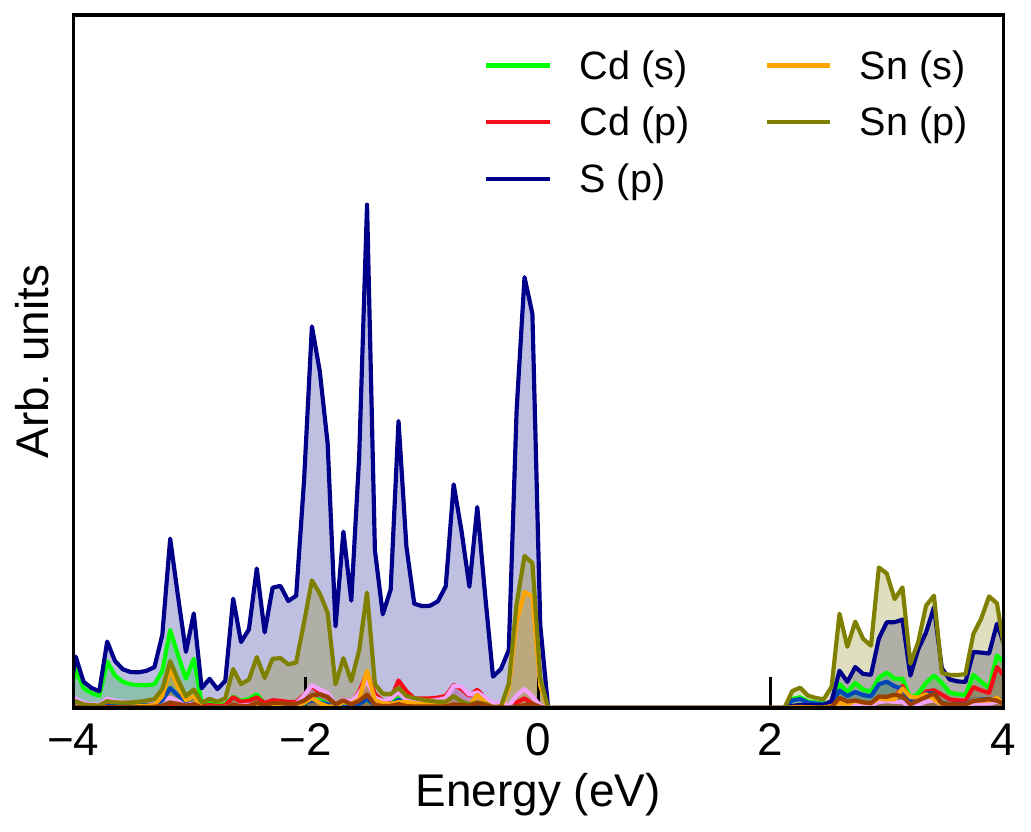}
\includegraphics[scale=0.3]{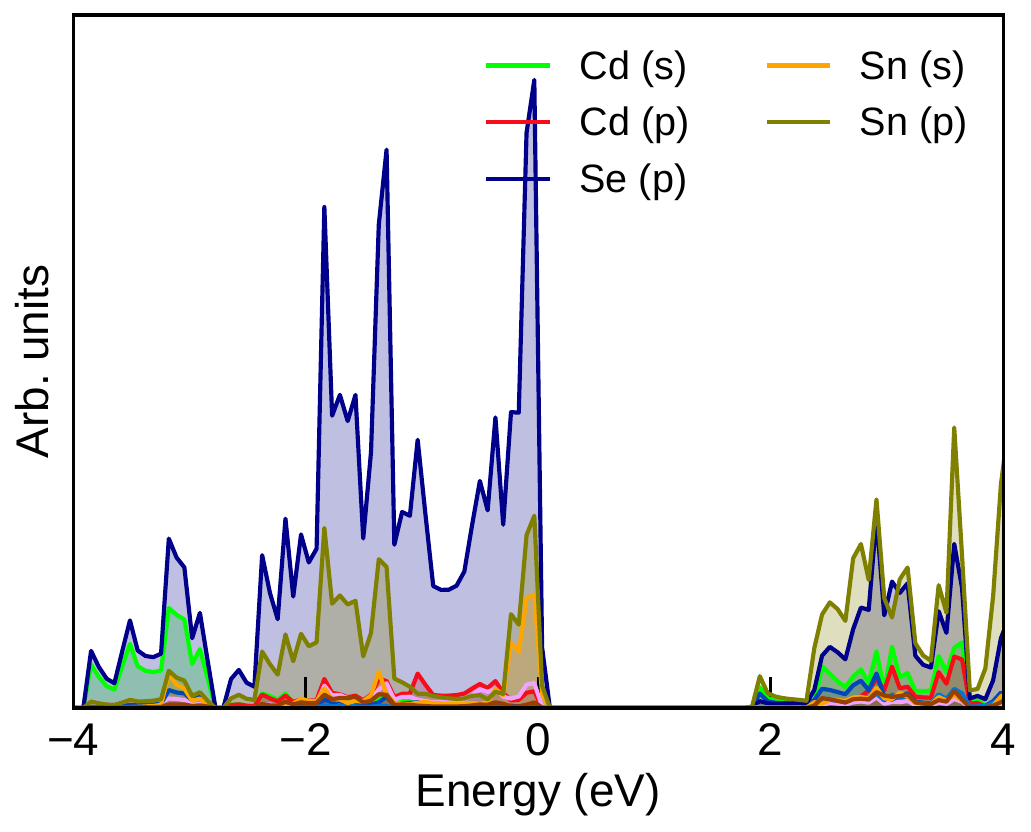}
\includegraphics[scale=0.3]{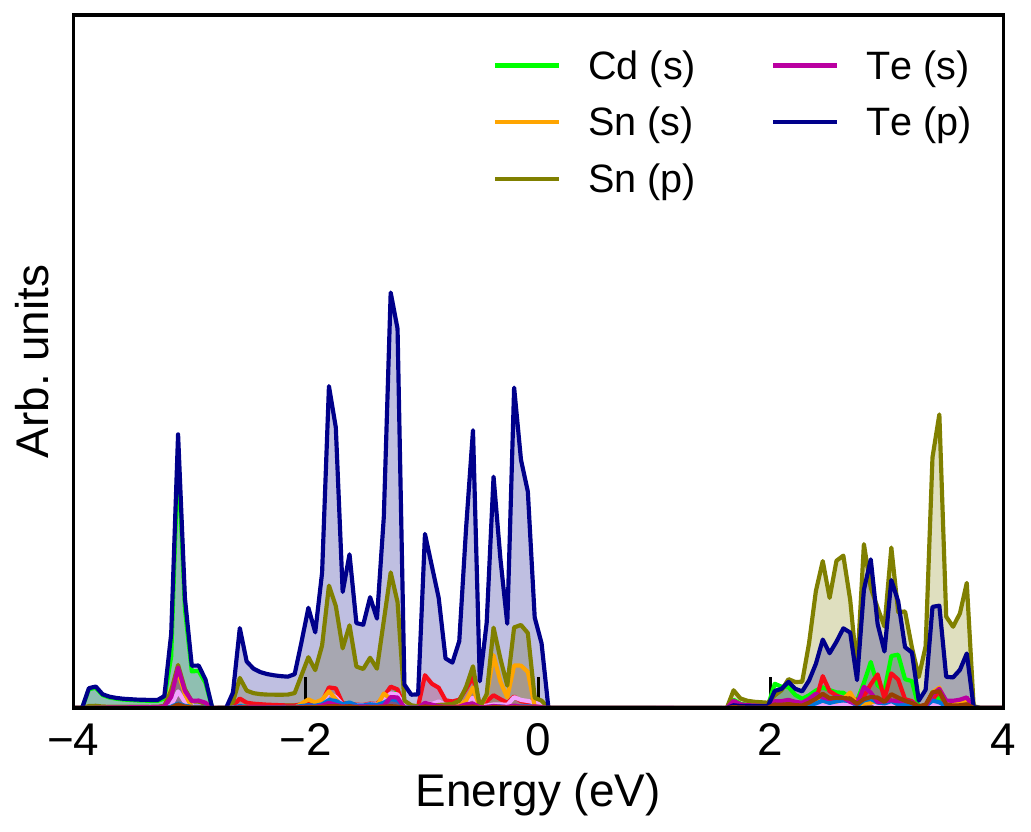}\\
\vspace{-0.08in}
 (a) \hskip 1.8in (b) \hskip 1.85in (c) \\
\vskip 0.05in
\includegraphics[scale=0.3]{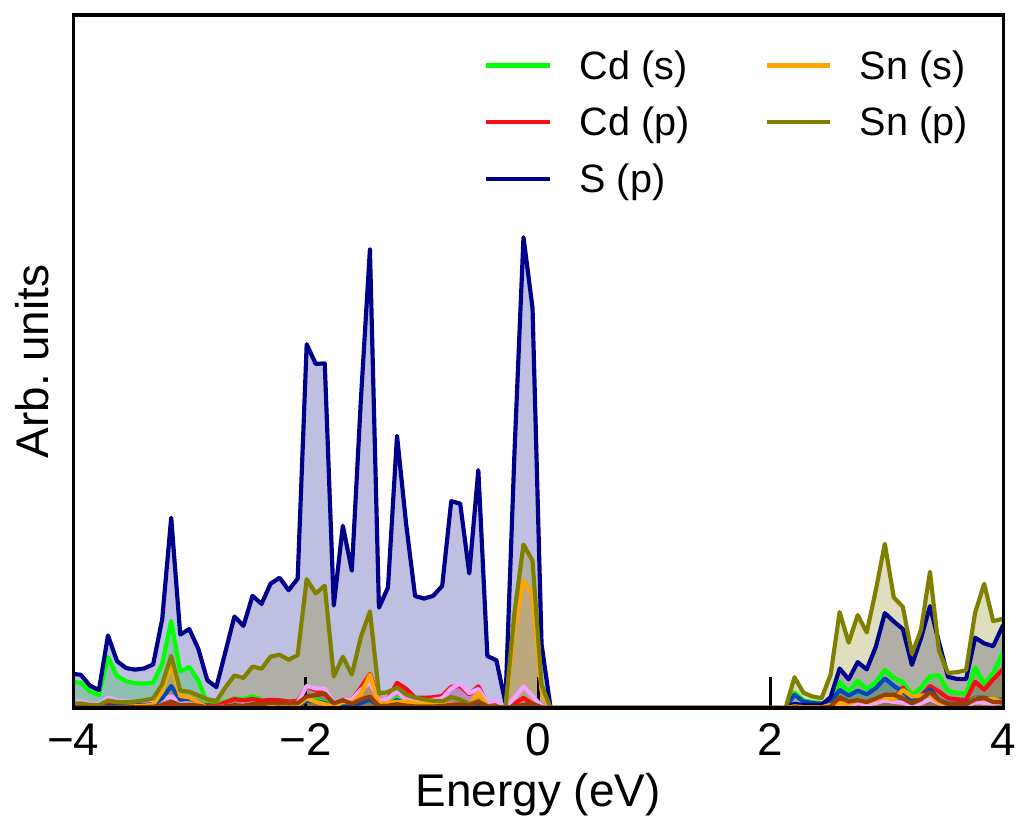}
\includegraphics[scale=0.3]{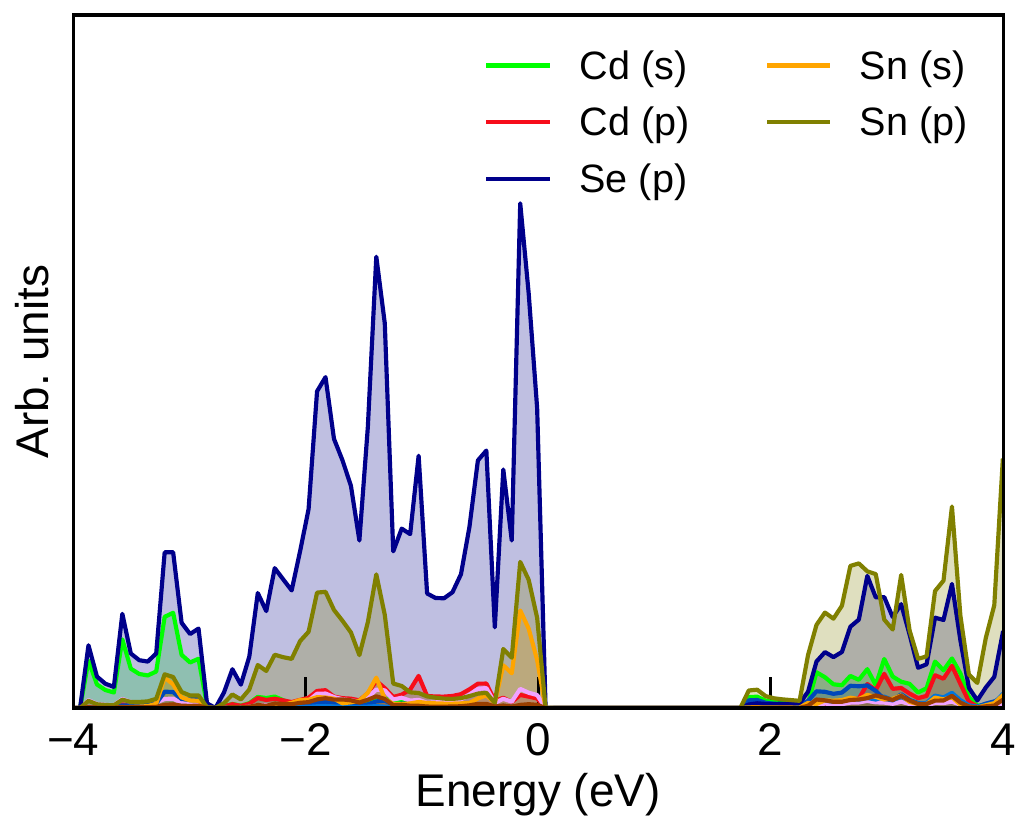}
\includegraphics[scale=0.3]{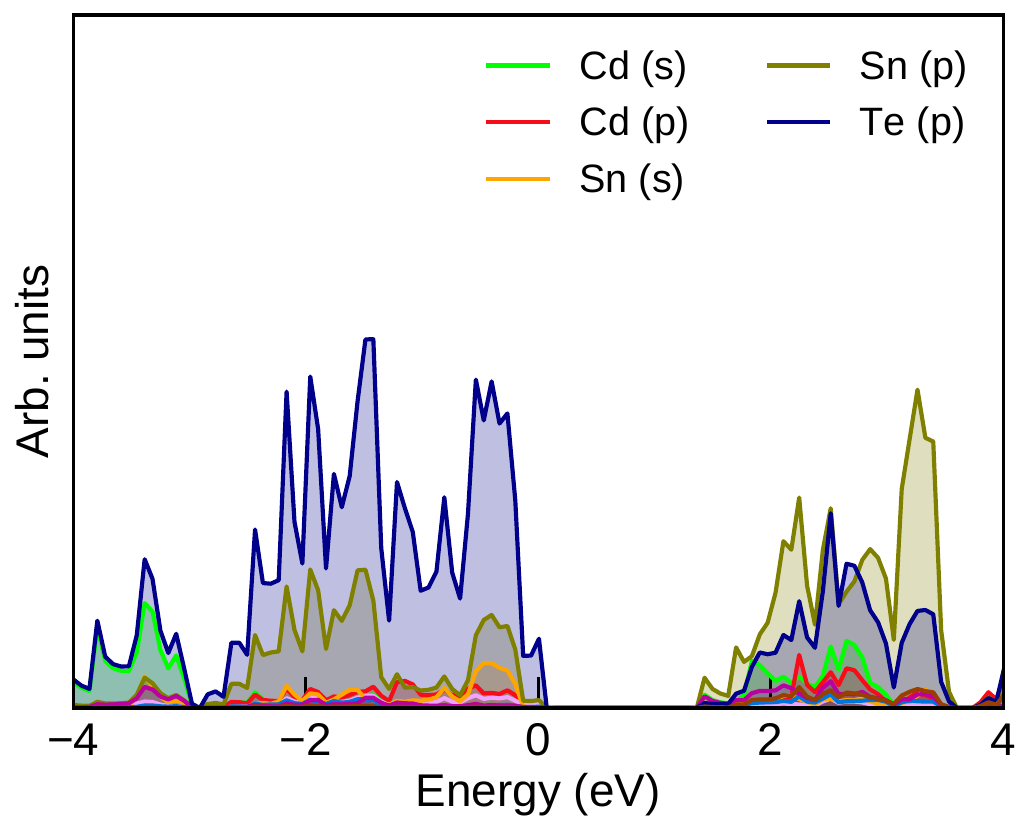}\\
\vspace{-0.08in}
 (d) \hskip 1.8in (e) \hskip 1.85in (f) \\

\vspace{-0.06in}
{\small{{\bf Figure S7.} (Color online) Projected density of states plots of Sn-doped CdX (X = S, Se and
Te) monolayers, without inclusion of SOC, are shown in (a), (b) and (c) respectively. 
In (d), (e) and (f), the same are shown with inclusion of SOC.}}
\label{dos-Sn-doped}
\end{figure*}

\begin{figure*}
\begin {centering}
\includegraphics[scale=1.07]{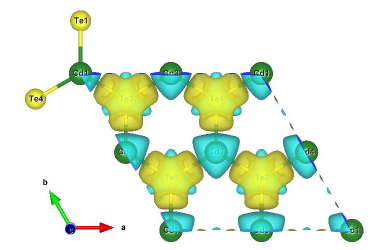}
\includegraphics[scale=1.07]{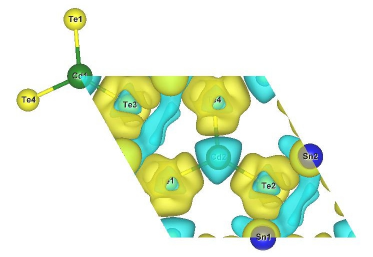}\\
\end{centering}
   (a) \hskip 2.50in (b)  

{\small{{\bf Figure S8.} (Color online) Difference charge density isosurface plots (for the same isovalue)  for
(a)~pristine CdTe monolayer and (b)~Sn-doped CdTe monolayer.
Yellow and cyan colors respectively depict electron accumulation and electron depletion.}}
\end{figure*}

In order to understand the effect of Sn-doping in CdX monolayers on the stability of the monolayers 
and on the changes in bonding, we have calculated the difference charge
density defined by Eq.~(\ref{charge})
\begin{equation}
 \rho_{diff}(\vec{r}) = \rho_{mono} (\vec{r}) -  \sum_{i}\rho_i(\vec{r}) 
 \label{charge}
\end{equation}
Here $\rho_{mono}(\vec{r})$ is the total charge density of the pristine or Sn-doped monolayer at point $\vec{r}$ and
$\rho_i(\vec{r})$ is the charge density of the $i^{th}$ constituent atom at point $\vec{r}$
calculated from single point  
calculations. The difference charge density isosurface plots for pristine and
Sn-doped CdTe monolayers
are shown for comparison in Figs.~S8~(a) and (b) respectively.  It is seen that for CdTe, each Cd atom has three
Te atoms as near-neighbors and vice versa. Each Cd atom donates equal charge to each Te neighbor and the
difference charge density has three-fold symmetry. On the other hand, in Sn-doped CdTe,
there is more charge build-up between Te and Sn atoms in comparison to Te and Cd atoms. Moreover,
there is additional charge depletion between the two Sn atoms which is absent in pristine 
structures between two Cd atoms.
This results in stronger binding between the atoms in Sn-doped monolayers. Thus, the Sn-doped 
monolayers are more stable than their corresponding pristine layers. This is reflected in the 
$E_{coh}^{CdX}$ and  $E_{coh}^{Sn-doped CdX}$ values as listed in Table~1 of the paper.

\begin{figure*}
\begin{centering}
 \includegraphics[scale=0.75]{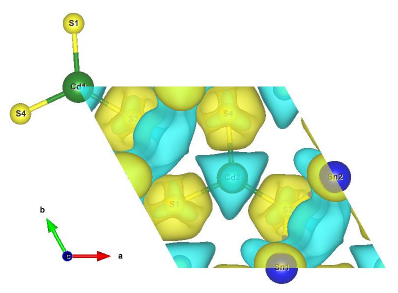}\hskip -0.07in\includegraphics[scale=0.75]
{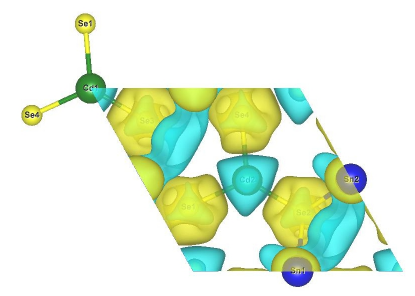}\hskip -0.07in\includegraphics[scale=0.8]{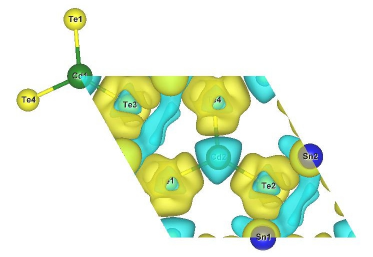}\\
\end{centering}
(a) \hskip 1.8in (b) \hskip 1.6in (c) 

{\small{{\bf Figure S9.} (Color online) Difference charge density isosurface plots (for the same isovalue)  of Sn-doped
CdX (X = S, Se and Te) monolayers 
with SOC. Yellow color indicates charge accumulation and cyan color depicts charge depletion regions.}}

\end{figure*}

Two Sn atoms share electrons with X2 and X3 atoms (please refer to Fig. 1(b) of the main text of 
the paper for the atomic labels) and X1 and X4 atoms share electrons with Cd atoms and electron 
enriched regions appear near X atoms. 
The difference charge density isosurface plots for Sn-doped CdX monolayers 
are shown in Fig.~S9~(a) (b) and (c)  for X = S, Se and Te respectively.
As we change the chalcogen from S to Se and then to Te,  with the increase in the size of the X atom, 
the charge transfer to the X atom decreases. Thus the covalency of the bonds increases as we move 
from S to Te, more so in Sn-doped monolayers.

To understand the origin of Zeeman-like splitting observed in the band structure plots, 
we have also calculated the net atomic charge distributions 
using the Bader method. The calculated net atomic charges are tabulated in Table~S3. 
Bader charge analysis reflects that Cd atoms donate electrons to X atoms and the 
atomic charges on Cd and X atoms in a particular pristine monolayer are 
same for each of the Cd and X atoms. Also, the bond lengths between each of the Cd and 
its three nearest-neighbor X atoms 
are equal.  In case of Sn-doped monolayers, atoms X2 and X3  have equal Bader charge value while
atoms X1 and X4 have equal Bader charge value which is different from that of atoms X2 and X3.
After doping Sn atoms at Cd sites, two Cd atoms along with the neighboring X1 and X4 chalcogen atoms
form a group, say Subset1, while two Sn atoms with two neighboring X2 and X3 chalcogen atoms form another group,
say Subset2. The atoms in Subset1 are all in the same plane while the Sn atoms in Subset2 lie
slightly out-of-plane. Subset1 has a net negative charge while Subset2 has a net positive charge as 
revealed from Table~S3. This little charge imbalance may act like an internal electric field which 
may cause the spin-splitting, in addition to SOC.

\begin{table*}[h!]
{\small{{\bf Table S3.} The calculated net atomic charges (in units of $e$, the electronic charge)
from Bader charge analysis of pristine and Sn-doped CdX monolayers.}}

\vspace{0.1in}
\centering{
\begin{tabular}{ | c | c | c | c | c | c | c | }
\hline
 &   \multicolumn{3}{c|}{Pristine CdX monolayer} &  \multicolumn{3}{c|}{Sn-doped CdX monolayer}\\
    \hline
 Charges on atoms $\downarrow$& ~~CdS~~  & ~CdSe~~ & ~CdTe~~ & ~~CdS~~ & ~CdSe~ & ~CdTe~~\\
\hline
Cd & 0.762 (0.86)$^a$ & 0.708 & 0.506 & 0.788 & 0.652 & 0.463\\
 \hline
X1 and X4  & -0.762 (-0.86)$^a$ &- 0.708&-0.506 & -0.817 & -0.679  & -0.489 \\
\hline
Subset1 & 0.00& 0.00& 0.00& -0.058 & -0.054 & -0.052\\
\hline
X2 and  X3   &-0.762 (-0.86)$^a$ &-0.708 & -0.506  & - 0.820 &  -0.693 &-0.508\\
\hline
 Sn  & - & - & - &0.849 & 0.720 & 0.534\\
 \hline
 Subset2 & - & - & - & 0.058 & 0.054 & 0.052\\
\hline
\end{tabular}}
\end{table*}
\vspace{-0.2in}
a : Ref.~\cite{pgarg}.

\begin{table*}
{\small{{\bf Table S4.} Orbital and spin-components of the atoms participating in the band splitting at K
and K$\textquotesingle$ for the highest valence band (VBM) and the band just below this valence band 
(BVB) for pristine monolayers. The (+) and (-) signs indicate the orientation of the spin component 
with respect to the specific ${\vec{k}}$-point.}}\\
\centering{
\begin{tabular}{| c | c | c | c  |}
\hline
 &   \multicolumn{3}{c|}{Pristine CdX monolayer} \\  
    \hline
  &  CdS & CdSe & CdTe\\
\hline
K&\multicolumn{3}{c|}{ }\\
\hline
VBM~&~S~~ $p_z[S_z(-)]$ &~Se~~ $p_z[S_z(-)]$ & ~Te~~ $p_z[S_z(-)]$ \\ 
 \hline
BVB~&~S~~ $p_z[S_z(+)]$ &~Se~~ $p_z[S_z(+)]$ &~Te~~ $p_z[S_z(+)]$   \\
 \hline
K$\textquotesingle$ &\multicolumn{3}{c|}{ }\\
 \hline
VBM~ &~S~~ $p_z[S_z(+)]$ & ~Se~~ $p_z[S_z(+)]$ & ~Te~~ $p_z[S_z(+)]$  \\
  \hline
BVB~&~S~~ $p_z[S_z(-)]$ &~Se~~ $p_z[S_z(-)]$ & ~Te~~ $p_z[S_z(-)]$   \\
   \hline
\end{tabular}
}
\end{table*}

\begin{table*}
{\small{{\bf Table S5.} Orbital and spin-components of the atoms participating in the band splitting at
K and K$\textquotesingle$ for the highest valence band (VBM) and the band just below this valence 
band (BVB) for Sn-doped monolayers. The (+) and (-) signs indicate the orientation of the spin component 
with respect to the specific ${\vec{k}}$-point.}}

\vspace{0.1in}

\begin{centering}
\begin{tabular}{|c |c |c|}
\hline
   &\multicolumn{2}{c|}{Sn-doped CdX monolayer} \\  
    \hline
  & \multicolumn{2}{c|}{CdS} \\ 
\hline
K&\multicolumn{2}{c|}{ }\\
\hline
VBM&S~$p_y[S_x(+)],p_z[S_x(+),S_y(-),S_z(-)]$&Sn~ $s[S_x(+),S_y(-)],p_z[S_x(+),S_y(-),S_z(-)]$ \\ 
 \hline
BVB&S~ $p_y[S_x(-)],p_z[S_x(-),S_y(+),S_z(+)]$ & Sn~ $s[S_x(-),S_y(+)],p_z[S_x(-),S_y(+),S_z(+)]$  \\
 \hline
 K$\textquotesingle$ &\multicolumn{2}{c|}{ }\\
 \hline
VBM &S~$p_y[S_x(-)],p_z[S_x(-),S_y(+),S_z(+)]$ &Sn~ $s[S_x(-),S_y(-)],p_z[S_x(-),S_y(+),S_z(+)$    \\
  \hline
BVB &S~$p_y[S_x (+)],pz[S_x(+),S_y(-),S_z(-)]$&Sn~$s [S_x(+),S_y(-)],p_z[S_x(+),S_y(_),S_z(-)$\\
\hline
 & \multicolumn{2}{c|}{CdSe} \\ 
\hline
K&\multicolumn{2}{c|}{ }\\
\hline
VBM &Se~$p_y[S_x(+),S_y(-)],p_z[S_x(+),S_y(-),S_z(-)]$&Sn~$s [S_x(+),S_y(-)],p_z[S_x(+),S_y(-),S_z(-)]$ \\ 
 \hline
 BVB &Se~$p_y[S_x(-),S_y(+)],p_z[S_x(-),S_y(+),S_z(+)]$ &Sn~$s [S_x(-),S_y(+)],p_z[S_x(-),S_y(+)]$  \\
 \hline
 K$\textquotesingle$ &\multicolumn{2}{c|}{ }\\
 \hline
VBM &Se~$p_y[S_x(-),S_y(+)],p_z[S_x(-),S_y(+),S_z(+)]$ &Sn~ $s[S_x(-),S_y(+)],p_z[S_x(-),S_y(+),S_z(+)] $  \\
  \hline
BVB &Se~$p_y[S_x(+),S_y(-)],pz[S_x(+),S_y(-),S_z(-)]$&Sn~$s [S_x(+),S_y(-)],p_z[S_x(+),S_y(-)]$\\
\hline
& \multicolumn{2}{c|}{CdTe} \\ 
\hline
K&\multicolumn{2}{c|}{ }\\
\hline
VBM &Te~$p_y[S_x(+),S_y(-)],p_z[S_x(+),S_y(-)]$&Sn~ $s[S_x(+),S_y(-)],p_z[S_x(+),S_y(-)]$ \\ 
 \hline
 BVB &Te~$p_z[S_x(-),S_y(+)]$ &Sn~ $s[S_x(-),S_y(+)],p_z[S_x(-),S_y(+),S_z(-)]$  \\
 \hline
 K$\textquotesingle$ &\multicolumn{2}{c|}{ }\\
 \hline
VBM &Te~$p_y[S_x(-),S_y(+)],p_z[S_x(-),S_y(+)]$ &Sn~$s [S_x(-),S_y(+)],p_z[S_x(-),S_y(+)]$     \\
 
  \hline
BVB &Te~$p_z [ S_x(+),S_y(-)]$&Sn~$s [ S_x(+),S_y(-)],p_z [ S_x(+),S_y(-),S_z(+)]$\\
\hline
\end{tabular}
\end{centering}
\end{table*}

\begin{figure*}

.\hskip 0.25in
\includegraphics[scale=0.57]{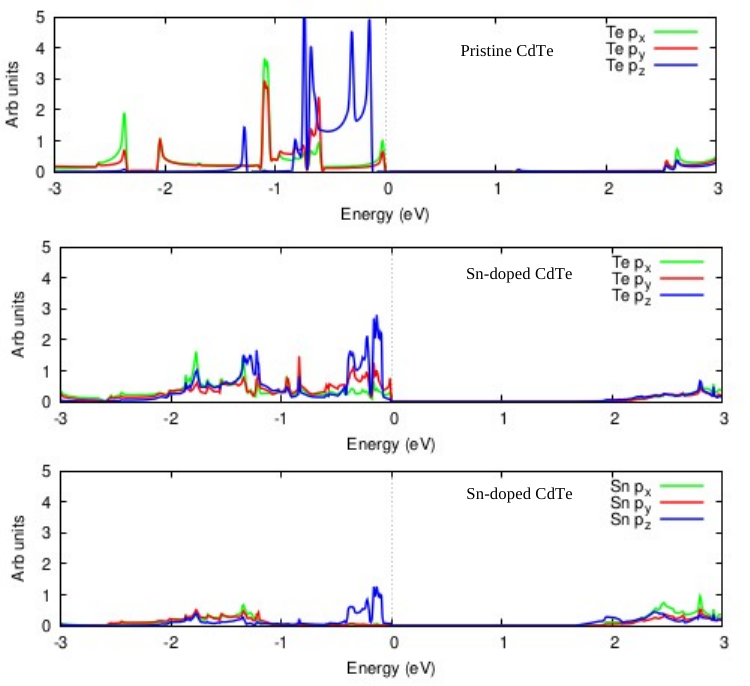}  \includegraphics[scale=0.58]{S_fig10_a.pdf}

\vskip -0.1in
 (a) \hskip 2.7in (b)

{\small{{\bf Figure S10.} (Color online) $p_x$, $p_y$ and $p_z$ site (atom) projected DOS for CdTe monolayer (a)~without and (b)~with inclusion of SOC.
The top panels in both the figures depict $p$-DOS for Te in pristine monolayer, middle panels depict 
$p$-DOS for Te in Sn-doped monolayers
and the bottom panels depict $p$-DOS for Sn in Sn-doped monolayers.}}
\end{figure*}
\newpage
\vskip -0.8in
\noindent

{\section{References}}

\bibliography{ref1}